\documentclass[referee]{raa}            
\usepackage{graphicx,times}             
\usepackage{natbib}
\usepackage{amssymb,amsmath}
\usepackage{textcomp}
\usepackage[]{caption} 
\bibpunct{(}{)}{;}{a}{}{,}
\usepackage[pagebackref=true]{hyperref}

\begin{document}

  \title{Astronomical Test with CMOS on the 60-cm Telescope at the Xinglong Observatory, NAOC
}

   \volnopage{Vol.0 (20xx) No.0, 000--000}      
   \setcounter{page}{1}          
   \author{Hai-Yang Mu  
      \inst{1,2}
    \and Zhou Fan
      \inst{1,2}
    \and Yi-Nan Zhu
      \inst{1}
    \and Yu Zhang 
      \inst{1}
   \and Hong Wu 
      \inst{1,2}
   }

   \institute{Key Laboratory of Optical Astronomy, National Astronomical Observatories, Chinese Academy of Sciences, Beijing 100101, People’s Republic of China; {\it zfan@nao.cas.cn; ynzhu@bao.ac.cn; yzhang@bao.ac.cn; hwu@bao.ac.cn; hymu@bao.ac.cn}\\
        \and
            School of Astronomy and Space Science, University of Chinese Academy of Sciences, Beijing 100049, People’s Republic of China\\
\vs\no
   {\small Received 20xx month day; accepted 20xx month day}}

\abstract{ This work shows details of an evaluation of an observational system comprising a CMOS detector, 60-cm telescope, and filter complement. The system's photometric precision and differential photometric precision, and extinction coefficients were assessed through observations of Supersky flat fields, open clusters, standard stars, and exoplanets. Photometry was precision achieved at the 0.02 mag level, while differential photometry of 0.004 mag precision. Extinction was found to be agreed with previous studies conducted at Xinglong Observatory.  Ultimately, the results demonstrate this observing system is capable of precision scientific observations with CCD across the optical wavelengths.
}

   \authorrunning{Hai-Yang Mu, Zhou Fan, Yi-Nan Zhu, Yu Zhang \& Hong Wu   }            
   \titlerunning{Astronomical Test with CMOS on the 60-cm Telescope at the Xinglong Observatory, NAOC}  

   \maketitle

\section{Introduction}           
\label{sect:intro}
\hspace{2em}
Since the application of charge-coupled devices (CCDs) (\citealt{Boyle+Smith+1993})to astronomical observation, there is a significant increase in observation methods. Later with developments in modern semiconductor technology, complementary metal-oxide-semiconductor (CMOS) sensors appeared. However, early CMOS technology was not suitable for scientific astronomical observation due to the high readout noise, and the lower dynamic range compared to CCD technology (\citealt{Gallaway+2016}). Furthermore, the independent readout of each pixel can potentially lead to non-uniformity in the readout levels between pixels.  Thus, early CMOS technology was primarily used for non-scientific astronomical imaging.

The advent of scientific CMOS image sensors led to the initial application of CMOS cameras in amateur astronomical observation (\citealt{Fossum+1997}; \citealt{Bonanno+etal+2003}). Advancements in CMOS manufacturing processes have been more and more  significant. Currently, only small difference exists between the readout noise of CMOS and CCDs.  Further, the dynamic range of CMOS sensors has expanded considerably. In addition, the emergence of back-illuminated CMOS chips has substantially increased quantum efficiency (\citealt{Bigas+etal+2006,Li+etal+2006}). The readout noise of CMOS approach to that of CCDs. As a result, the difference between the performance of CCD and CMOS image sensors continues to diminish. 

The CMOS applications are popularly used by amateur astronomers, and CCD is still it is widely used astronomical professional observation. However, the CMOS has been used in some sky survey projects, such as the Argus Optical Array(\citealt{Law+etal+2022}), the Large Array Survey Telescope(\citealt{Ofek+etal+2023}), or the next-generation telescope of the ATLAS project(\citealt{Tonry+etal+2018}), which will be installed at the ATLAS-Teide observatory on the Teide mountain(\citealt{Licandro+etal+2023}). Some of them use CMOS because of its low cost and ability of wide field-of-view and high signal-to-noise ratio measurements. In fact, there are also differences in performance between different brands of CMOS. A considerable amount of testing has been done on CMOS(\citealt{Qiu+etal+2021,Karpov+etal+2020,Alarcon+etal+2023}), which has revealed the following unique problems such as the Salt \& Pepper noise that are not seen with CCDs. However, these tests are solely on performance tests, and the astronomical tests of CMOS, such as photometric accuracy and extinction coefficient is rare.

In this article, we test the CMOS product the SONY IMX455 chip, back-illuminated. In section 2, the observation system is introduced. In section 3, the processing of the data reduction is described. In section 4, the results and analysis astronomical observations are introduced. The section 5 present the summary and conclusions.

\section{Observations}
\label{sect:Obs}

\subsection{Observation System}

\hspace{2em}The observation system used in this article includes telescope, CMOS camera, and filters.

The telescope used in this observation is the 60-cm telescope at the Xinglong Observatory of the National Astronomical Observatories of China (NAOC). The telescope has the following parameters due to the official observatory website\footnote{Reference Sites:\url{http://www.xinglong-naoc.cn/html/gcyq/60/detail-28.html}}:
\begin{itemize}
\item Aperture: 60 cm
\item Field of view: $18' \times 18'$
\item Focal ratio of primary mirror: $F/3.28$
\item Corrected focal ratio of primary focus: $F/4.23$
\end{itemize}

The 60-cm telescope was first light in 1964 as an technical test telescope for the 2.16-m telescope at Xinglong Observatory. It was fully commissioned in 1968 after serious of upgrades. Since 1974, the telescope has been used for photometric observations of variable stars such as eclipsing binaries and pulsating stars. From 1995 to 2000 it was used primarily for a supernova search survey. Since 2000 it has been focused on variable star observations.

The IMX455 \footnote{Reference Sites: \url{https://www.sony-semicon.com/files/62/pdf/p-13\_IMX455AQK\_BQK\_ALK\_Flyer.pdf}}(\citealt{Alarcon+etal+2023}) is a back-illuminated CMOS sensor, which is installed in ZWO ASI6200MM PRO camera in this observation system. Table \ref{tab:telsyr} shows the part of the parameters of the camera, and detailed information are referred to the official website\footnote{Reference Sites: \url{https://i.cmoscool.com/zwo-website/manuals/ASI6200MM\%20Pro_Manual_CN_V1.0.pdf}}. We can see that the dark current noise is only 0.003$e^-$/s/pix at 0 degrees Celsius,which means that a 5 minute exposure would result in only 0.9$e^-$ dark current noise, which is negligible. Pixel size is 3.76$\mu m$. Pixel scale is 0.306 arcsec/pixel for our testing system.
\begin{table}[htbp]
\begin{center}
\caption[]{Parameters of the CMOS camera}	  \label{tab:telsyr}
 \begin{tabular}{ll}
  \hline\noalign{\smallskip}
Features & Specifications      \\
  \hline\noalign{\smallskip}
  Sensor & SONY IMX455\\
Pixel number & $9576 \times \ 6833$ \\
Max full frames rate & 2fps\\
Pixel size & $3.76 \mu m \times \ 3.76\mu m$\\
Imageing area & $36 mm \times \ 24 mm$ \\
A/D conversion & 16bit\\
Full well & $51400e^-$\\
Readout noise & $1.2\sim 3.5e^-$\\
Dark current noise & 0.003$e^-$/s/pix at $0^\circ C$\\
Shutter Type & rolling shutter\\
Fill factor & 100\%\\
Operating temperature & Below ambient temperature $30^\circ C\sim 35^\circ C$\\
    \noalign{\smallskip}\hline
\end{tabular}
\end{center}
\end{table}
We used the Johnson-Bessell filter system(\citealt{Bessell+1990}). Only $BVRI$ are applied, and their corresponding central wavelengths are shown in the Table \ref{tab:filter}.\\
\begin{table}[htbp]
\begin{center}
\caption[]{Center wavelengths corresponding to $BVRI$ in Johnson-Bessell filter system}		\label{tab:filter}
 \begin{tabular}{ccccc}
  \hline\noalign{\smallskip}
$\lambda_c (nm)$ &  $B$      & $V$ & $R$ & $I$ \\
  \hline\noalign{\smallskip}
$ Johnson $ & 440 & 550 & 640 & 806$(Bessell)$\\
  \noalign{\smallskip}\hline
	\end{tabular}
\end{center}
\end{table}

\subsection{Observation}
\subsubsection{Observation plan for tests}

\hspace{2em}
In this article, the tests mainly include: Supersky flat field, accuracy of photometry, accuracy of differential photometry, and calculation of extinction coefficients during CMOS observations.
\\\\
\textit{Supersky flat field}

The primary goal of capturing a Supersky flat field is to compare it to dusk and dawn flat fields. This comparison allows one to assess differences between dusk and dawn sky illumination patterns. A Supersky flat field is obtained by observing low stellar density fields. A fixed exposure time was adopted, with the telescope being offset slightly between individual exposures, yielding approximately 300 images acquired across three bands. 
\\\\
\textit{Open clusters}

To measure the photometric accuracy, we observed four open clusters, NGC6913, NGC7243, NGC6811 and NGC744, with consideration for time constraints on individual exposures. Open clusters were selected based on having an optimal density of stars within the field-of-view to balance sufficient numbers with avoiding excessive crowding. Additionally, open clusters contain a large population of variable stars and binary systems, as noted by \cite{Abazajian+etal+2009}. These unique features of open clusters informed our choice to utilize them in our calculations. Table \ref{xw} lists the specific targets of the photographed open clusters in our observing program. Open clusters serve as ideal targets to test photometric accuracy through calibration to literature values.
\\\\
\textit{Standard stars}

The extinction coefficients were computed from observations of standard stars. The method is monitoring standard stars throughout all night to from rise to set. Photometric standard stars are precisely determined flux measurements across various photometric systems \citep{Oke+etal+1983,Landolt+2007,Landolt+2013}. Through measurements using a CCD camera or photometer, the brightness or flux of another object can be determined by comparison to the standards. By tracking standard star magnitudes versus airmass, the extinction coefficient was derived. In this work, standard stars were from Landolt's catalog\citep{Landolt+2007,Landolt+2013}. Table \ref{xw} lists the standard stars selected for our observations.
\\\\
\textit{Exoplanet}

To measure the differential photometric precision, we observe two exoplanets, Table~\ref{xw} lists the two observed targets: HAT-P-32 b and WASP-33 b. The approach involved capturing single-band V exposure frames during the transit. Both targets were monitored continuously from one hour before until one hour after the transit event of exoplanet. 

\begin{table}[htbp]
	\begin{center}
		\caption{Observe the information of the object.}		\label{xw}
		\begin{tabular}{cccccccc}
		\hline\noalign{\smallskip}
		Name & Ra(J2000) & Dec(J2000) & Vmag & Deep & Duration/min & Type & Purpose\\
			\hline\noalign{\smallskip}
			 NGC6913 & 01 50 41.71 & +21 45 35.89 & 6.6 &-- &--& Open \ cluster & photometric accuracy\\
			 NGC7243 & 22 15 9.12  & +49 49 48.00 & 6.4 &-- & --& Open \ cluster & photometric accuracy\\
			 NGC6811 &19 37 21.60 & +46 22 40.80 &7.74 &-- &--& Open \ cluster & photometric accuracy\\
			 NGC744 & 01 58 36.480 & +55 28 22.80 &8.38 &-- &--& Open \ cluster & photometric accuracy\\
			 SA 20-43 &  00 45 42.45 & +45 35 15.43	& 10.4&-- &--& Standard \ star & extinction coefficients\\
			 HD 165434 & 02 37 50.79 & -13 07 45.29 & 7.71&-- &--& Standard \ star & extinction coefficients\\
			HAT-P-32 b & 02 04 10.24 & +46 41 16.8 & 11.29 & 0.0244 & 186.5& Exoplant & differential photometric accuracy\\
			WASP-33 b & 02 26 51.08 & +37 33 02.5 & 8.3 & 0.0151 & 163 & Exoplant & differential photometric accuracy\\
			\noalign{\smallskip}\hline
		\end{tabular}
	\end{center}
\end{table}
\subsubsection{Observation}
\hspace{2em}Due to the limit of mechanical tracking the 60-cm telescope, the maximum observation time for a single target is 30 seconds. For the camera settings, the GAIN parameter is set to 100 which refers to the gain value of $0.25e^-/ADU$, and the readout noise value is $1.5e^-$.  The camera was cooled to -10 degrees Celsius. 

Our observation is in Table \ref{tab:log}. The relevant information includes: Primary Observation Targets,  Observational time, Band, Exposure, and Frame. Our main observation period is from September 23 through September 30, 2022. Prior to each night, the imaging quality was assessed and the optical focus adjusted to optimize image quality. Each open cluster was observed sequentially in 4 filters, 80 images per band. Standard stars were monitored throughout entire nights in all 4 filters. Exoplanet targets were observed in V band alone for the whole night. The observation strategy employed is different between target: HAT-P-32 b, which was in normal focus, and WASP-33 b, which was defocused.
\\
\begin{table}[htbp]
\centering
\caption[]{The observation of the Log table, basically have Primary Observation Targets,  Observational time, Band, Exposure, and Frame.}		\label{tab:log}
\resizebox{\textwidth}{!}{ 
\begin{tabular}{ccccc}
\hline\noalign{\smallskip}
Primary Observation Targets &  Observational \ time  & Band & Exposure \ time  & Frame  \\
 &  (YYYY/MM/DD) &  & (s)  &  \\
  \hline\noalign{\smallskip}
 NGC6913  & 2022/09/24 & B & 5  &  80 \\
 &  & V & 5 & 80  \\
 &  & R & 5 & 80  \\
 &  & I & 5 & 80  \\
 NGC7243  & 2022/09/24 & B & 5  & 80 \\
 &  & V & 5 & 80  \\
 &  & R & 5 & 80  \\
 &  & I & 5 & 80  \\
 HAT-P-32 b  & 2022/09/26 & V & 30 & 541  \\
 WASP-33 b  & 2022/09/27 & V & 20 & 548 \\
 SA 20-43 & 2022/09/28 & B & 20 & 282  \\
 &  & V & 20 & 282  \\
 &  & R & 10 & 282  \\
 &  & I & 20 & 282  \\
 Supersky flat field  & 2022/09/29 & B & 20 & 100 \\
 &  & V & 20 & 284  \\
 &  & R & 20 & 100  \\
 NGC6811  & 2022/09/30 & B & 30 & 80 \\
 &  & V & 30 & 80 \\
 &  & R & 15 & 80 \\
 &  & I & 15 & 80  \\
 NGC744  & 2022/09/30 & B & 25 & 80 \\
 &  & V & 25 & 80 \\
 &  & R & 25 & 80 \\
 &  & I & 25 & 80  \\
 HD 165434 & 2023/06/21 & B & 4 & 532 \\
 &  & V & 4 & 532  \\
 &  & R & 4 & 532 \\
 &  & I & 16 & 532 \\
\noalign{\smallskip}\hline
\end{tabular}}
\end{table}

\section{Data reduction}
\label{sect:data}

\subsection{Image trimming}
\hspace{2em}As described in section 2, the approximate field-of-view of the 60-cm telescope 18$\arcmin$  $\times$ 18$\arcmin$ . However, during observations it was found that the full field captured by the camera subtended roughly 32$\arcmin$  $\times$ 48$\arcmin$. Therefore, trimming of the images was required to extract the relevant field-of-view for analysis. To obtain the best illuminated of the telescope.
Only a subsection of the full CMOS frame corresponding to the area illuminated by the telescope optics was retained for photometry.

\begin{figure}[htbp]
  \centering
  \includegraphics[width=\textwidth]{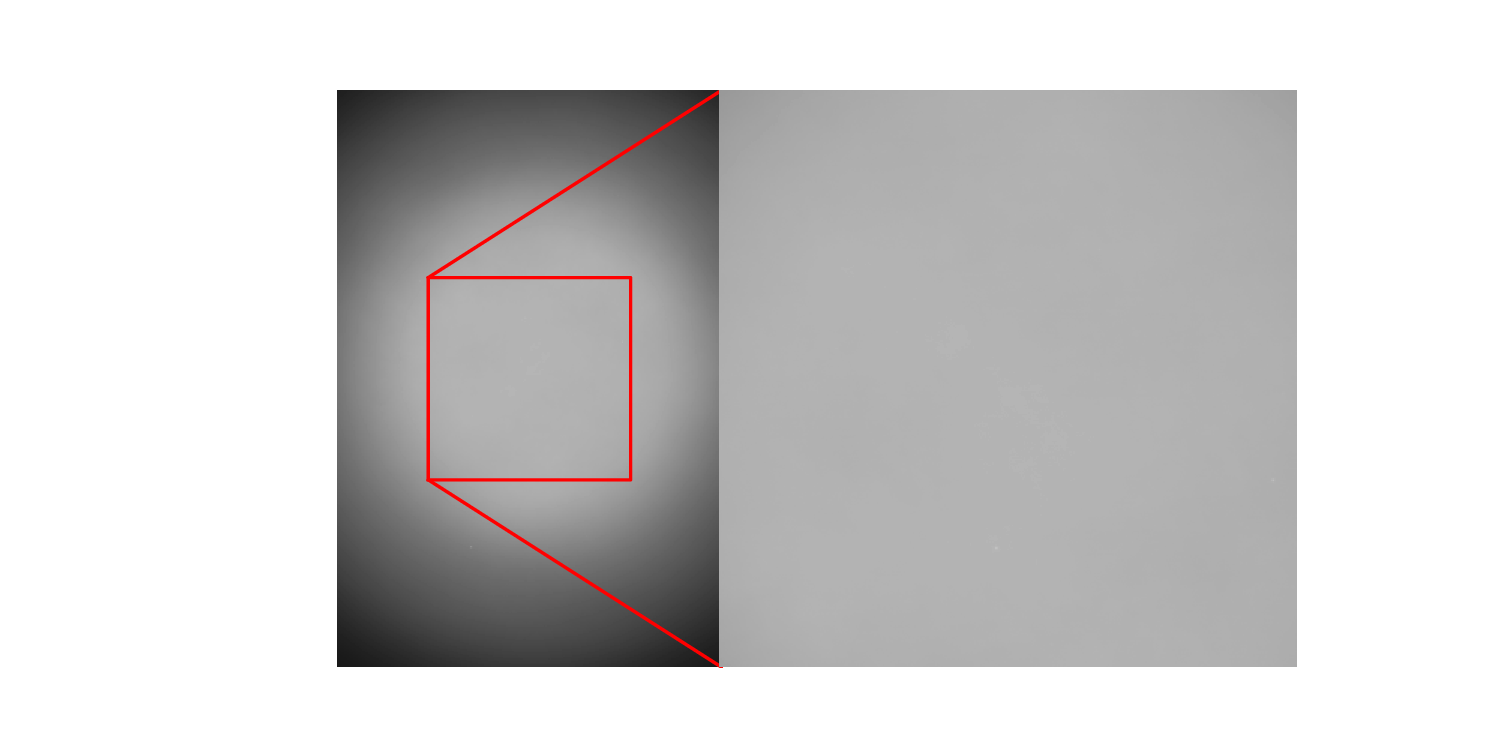}
  \caption{Image data trim, the right panel is before trim, the left panel is after trim.}
  \label{fig:trim}
\end{figure}

After the trimming, the image data size is changed to 3354 $\times$ 3354 $pixel^2$, and its field of view is changed to 17.05 $\arcmin$ $\times$ 17.05 $\arcmin$. 
The pre-trimmed and post-trimmed planar fields are shown in Figure \ref{fig:trim}, where the left is the pre-trimmed planar field, and the box in the figure is the trimmed range. 
The right panel is the trimmed flat field. All bias frames, flat fields, and science target images were trimmed to further processing and analysis. 
\subsection{Bias}
\hspace{2em}
This section aims to evaluate the stability of CMOS bias frames and investigate any temporal variations Before and after observation with CMOS. Bias images was taken at dusk and dawn on different nights. 
Examination of the biases indicates the background stabilizes at approximately 510 ADU.
First we combined bias frames using median. The median value was retained to mitigate the effects of salt and pepper noise (\citealt{Alarcon+etal+2023}). 
Differences between biases were analyzed to highlight any temporal variations. 
Table~\ref{bias1} shows the differences between the mean of the combined dusk bias and the combined dawn bias each day.
Table~\ref{bias2} shows the differences between the mean of the 24th combined dusk and the other's day combined dusk biases.

 \begin{table}[htbp]
    \caption{Bias compare, the difference value is the numerical difference between the highest peak and the lowest valley in the resulting image post the subtraction of the two bias images.}  \label{bias}
    \centering
    \begin{subtable}[t]{0.45\linewidth}
   	\centering
     \subcaption{Dawn bias versus Dusk bias,Dusk bias subtract Dawn bias }
      \label{bias1}
    \begin{tabular}{cc}
		\hline\noalign{\smallskip}
		Date (YYYYMMDD) & Difference value (ADU)\\
		\hline\noalign{\smallskip}
			20220924 & -0.09\\
			20220925 & -0.048\\
			20220926 & -0.072\\
			20220927 & -0.054\\
			20220928 & -0.074\\
			20220929 & -0.007\\
		\noalign{\smallskip}\hline
        \end{tabular}
        \end{subtable}
    \hfill
    \begin{subtable}[t]{0.45\linewidth}
    \centering
    \subcaption{With 24 Dusk bias, others day's dusk bias subtract 24 Dusk bias }
        \label{bias2}
	\begin{tabular}{cc}
	\hline\noalign{\smallskip}
		Date (YYYYMMDD) & Difference value (ADU) \\
		\hline\noalign{\smallskip}
			20220925 & 0.02\\
			20220926 & 0.044\\
			20220927 & 0.044\\
			20220928 & 0.063\\
			20220929 & 0.045\\
		 \noalign{\smallskip}\hline
        \end{tabular}
        \end{subtable}
\end{table}
These comparisons demonstrate the bias stability over the whole observation nights. Thus, we conclude the CMOS detector operation was stable throughout the observation period. During pre-processing, the dusk bias frame was subtracted from both the raw science images and flat fields at each night. 
Overall, the characterized bias stability supports reuse of a single master bias constructed from dusk frames only, without need to consider time-variable corrections over the observation run.

\subsection{Flat Field}
\hspace{2em}Flat field is essential for processing observational data not only it corrects for the unevenness of the illumination sky, but also correct difference between the amplifiers of each pixel. 

Table \ref{flat1} presents the outcomes of calculating the ratio of the median combined flat field captured during dusk versus that obtained during dawn for each day. 
This ratio represents the variation between the maximum and minimum of the images of the two flat fields after normalizing, and smoothing via the median filter. 
This metric provides a measurement of the illumination uniformity achieved across each band flat. 
Comparing the values among the different dusk and dawn flats indicates any variation introduced by changing illumination conditions at dusk versus dawn. 
The Table \ref{flat1} illustrate the comparison within each band, revealing that the median difference within 1\% in the B, V, and R bands, and within 2\% in the I band. 

 Table \ref{flat2} presents results from dividing the median value of dusk flat field for each night during of September 24-29. 
 These values indicate the level of consistency between daily flats during the observation period. 
 As expected, flat fields agree to 1\% in all bands. 
 This consistency validates combined master flat field better than 1\%, at this level it will not significantly impact science photometry. 
 The temporal stability of the flats supports their use for corrections throughout the observation window without introducing spurious variations.

\begin{table}[htbp]
\begin{center}
\caption[]{Flat comparison table for different dusk and dawn. The table's filter presents the comparison result of each band. which is the result of the difference the Peak-to-Valley value after dividing the two smoothed images.}\label{flat1}
		\begin{tabular}{ccccc}
		\hline\noalign{\smallskip}
		{Date (YYYYMMDD)/Filter} & {B} $\%$  & {V} $\%$ & {R} $\%$ & {I} $\%$ \\
			\hline\noalign{\smallskip}
			20220924 & 0.5 & 0.4 & 0.4 & 0.9\\
			20220925 & 0.5 & 0.4 & 0.3 & 1.2\\
			20220926 & 0.2 & 0.3 & 0.1 & 1.4\\
			20220927 & 0.3 & 0.3 & 7 & 1.5\\
			20220928 & 0.4 & 0.5 & 0.6 & 1.7\\
			20220929 & 0.9 & 0.9 & 0.9 & UNOBSERVED\\
			\hline\noalign{\smallskip}
		\end{tabular}
	\end{center}
\end{table}
\begin{table}[htbp]
	\begin{center}
		\caption{The rest of the day and 24 dusk flat comparison table. The table's filter presents the comparison result of each band. which is the result of the difference the Peak-to-Valley value by dividing the two smoothed images.}		\label{flat2}
		\begin{tabular}{ccccc}
		\hline\noalign{\smallskip}
		{Date (YYYYMMDD)/Filter} & {B} $\%$  & {V} $\%$ & {R} $\%$ & {I} $\%$ \\
			\hline\noalign{\smallskip}
			20220925 & 0.8 & 0.4 & 0.5 & 0.3\\
			20220926 & 0.6 & 0.3 & 0.5 & 0.3\\
			20220927 & 0.7 & 0.6 & 0.4 & 0.4\\
			20220928 & 0.6 & 0.5 & 0.6 & 0.3\\
			20220929 & 0.6 & 0.8 & 0.7 & UNOBSERVED\\
			\hline\noalign{\smallskip}
		\end{tabular}
	\end{center}
\end{table}

Therefore, we took the image that had the bias correction and dusk flat field to correction obtain the scientific images. 

\subsection{Photometry}
\hspace{2em} We used source-extractor tools for photometric. 
Source-extractor’s automatic aperture photometry routine derives from Kron’s “first moment” algorithm(\citealt{Kron+1980}). 
Details see source-extractor's manual(\citealt{Bertin+Arnouts+1996})\footnote{Reference Sites: \url{https://sextractor.readthedocs.io/en/latest/Introduction.html}}. 
In Table \ref{tab:sex_p}, we present some  parameters of the source-extractor. DETECT\_THRESH represents the threshold for star detection, for which we have opted for the default value. PHOT\_AUTOPARAMS denotes the parameters for automatic aperture photometry, for which we have also chosen the default settings. Additionally, BACKPHOTO\_TYPE specifies the background calculation method, with our selection being LOCAL to compute the flux error using local background estimation. The formula used by SE to calculate the flux error.
\begin{equation}
    Fluxerr=\sqrt{\sum_{i\in A }(\sigma_i^2 + \frac{p_i}{g_i})}
    \label{eq:fluxerr}
\end{equation}

Equation\ref{eq:fluxerr}:where $A$ is the set of pixels defining the photometric aperture, and $\sigma_i$ respectively the standard deviation of noise (in ADU) estimated from the local background, $p_i$ the measurement image pixel value subtracted from the background, and $g_i$ the effective detector gain in $e^{\text{-}}/ADU$ at pixel $i$. Note that this error estimate provides a lower limit of the true uncertainty, as it only takes into account photon and detector noise.

With aperture photometry and target extraction we can obtain the light curve of the target(\citealt{Huang+Xiao+Yuan+2022}). However, before we do this, we need to correct the exposure time. And then to correct the exposure time t and the magnitude $m_{phot}$ obtained from the photometry, to determine the ultimate instrumental magnitude. \\
\begin{equation}
    m_{inst}=m_{phot}+2.5\log_{10}(t)
    \label{eq:time}
\end{equation}
\begin{table}[htbp]
	\begin{center}
		\caption{source-extractor parameter}		\label{tab:sex_p}
		\begin{tabular}{cc}
		\hline\noalign{\smallskip}
		{Parameter} & {Specifications}\\
		\hline\noalign{\smallskip}
			DETECT\_MINAREA & 50 \\
			DETECT\_THRESH & 1.5\\
			PHOT\_AUTOPARAMS & 2.5, 3.5  \\
			SATUR\_LEVEL & 65525.0 \\
			MAG\_ZEROPOINT & 25 \\
			GAIN & 0.25 \\
			PIXEL\_SCALE & 0.306\\
			BACKPHOTO\_TYPE & LOCAL\\
			\hline\noalign{\smallskip}
		\end{tabular}
	\end{center}
\end{table}
\subsection{System conversion}
\hspace{2em}After the exposure time correction, we carry out the calibration and filter system correction. 
We introduce the reference catalog - Gaia's synthetic photometry(Gaia-SP)(\citealt{Gaia Collaboration+etal+2023}) and Gaia DR3(\citealt{Gaia Collaboration+etal+2023}).  
Here we applied Equation\ref{fum:mian}, to make corrections, and according to different situations to transform the formula, to simplify the process.

	\begin{equation}
	\begin{aligned}
	    m_{B,inst}=m_{B,Gaia-sp}+c_{B}+k_{1,B}\cdot X + k_{2,B} \cdot (m_{B,Gaia-sp}-m_{V,Gaia-sp})\\
		m_{V,inst}=m_{V,Gaia-sp}+c_{V}+k_{1,V}\cdot X + k_{2,V} \cdot (m_{B,Gaia-sp}-m_{V,Gaia-sp})\\
		m_{R,inst}=m_{R,Gaia-sp}+c_{R}+k_{1,R}\cdot X + k_{2,R} \cdot (m_{V,Gaia-sp}-m_{R,Gaia-sp})\\
		m_{I,inst}=m_{I,Gaia-sp}+c_{I}+k_{1,I}\cdot X + k_{2,I} \cdot (m_{V,Gaia-sp}-m_{I,Gaia-sp})
		\end{aligned}
		\label{fum:mian}
	\end{equation}
\hspace{2em}Where $m_{inst}$is the instrumental magnitude, $m_{Gaia-sp}$ is the magnitude with reference to GAIA-SP(\citet{Gaia Collaboration+etal+2023}) catalog, X is airmass, $k_1$ is the extinction coefficient corresponding to a certain band, $k_2$ is the color coefficient corresponding to the two systems, C is the zero point between the two system. 

Flux calibrations are usually calibrated using the field star maps are photographed and processed. The magnitude calibrations used here are those that convert the instrumental magnitude to the true magnitude, and here we use the Vega system. Because of the high accuracy Gaia DR3 data, we have selected Gaia DR3 as the reference magnitude.

\section{Results and Analysis}
\label{sect:analysis}

\subsection{Supersky flat field}

\begin{figure}
  \centering
   \includegraphics[width=80mm]{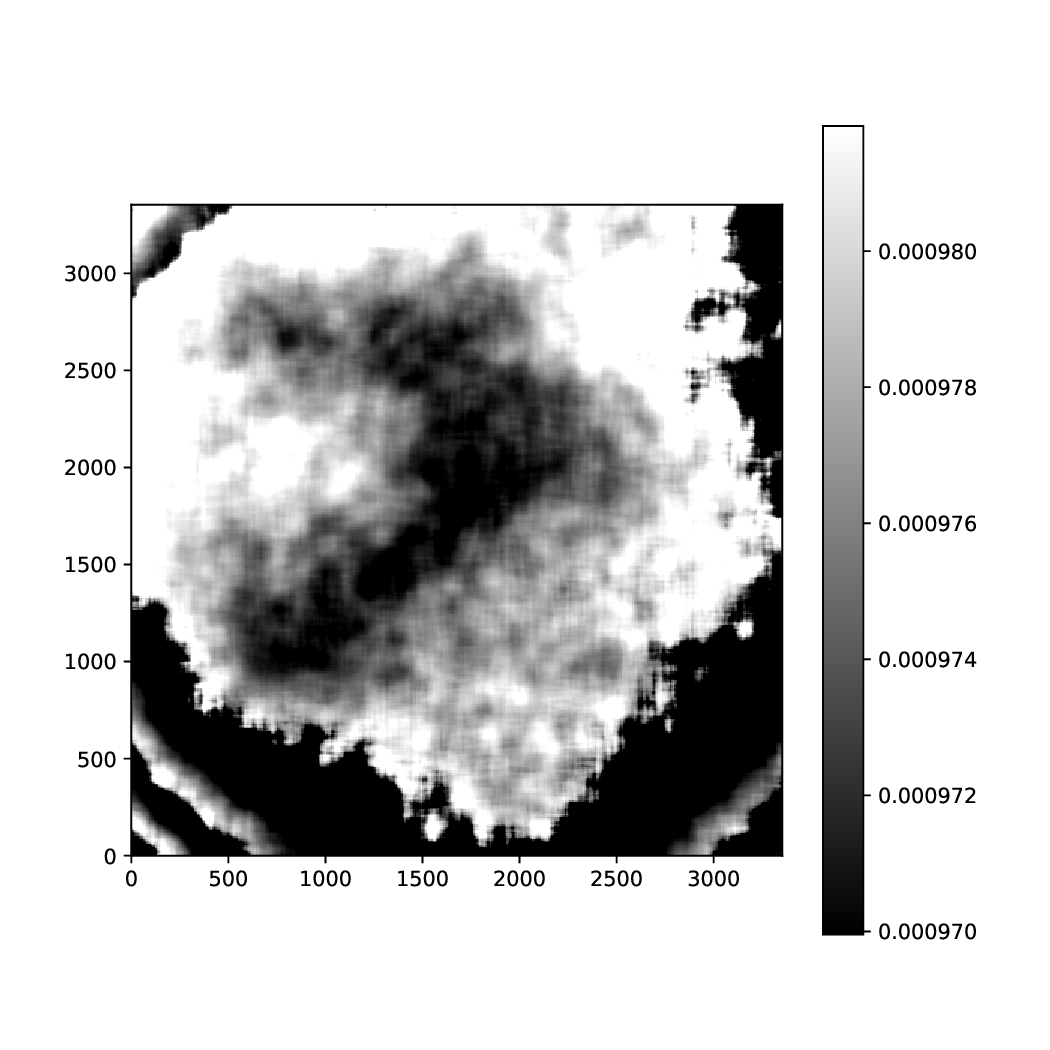}
	 \caption{\label{fig:supersky-flat}{\small This image is the division of Supersky field flat and the dusk and dawn flat field, and then the smoothed image with the box size of smooth is 100pixel.}}
\end{figure}

\hspace{2em}
This comparison evaluates the illumination differences Supersky flat field versus flat field from dusk and dawn. The comparison revealing discrepancies of sky background. 
Figure \ref{fig:supersky-flat} displays the image after dividing the Supersky flat field and the dusk and dawn flat field, followed by smoothing the image to a size of 100. Table \ref{tab:sp1} is obtained by calculating the difference the Peak-to-Valley value in this figure. It presents results comparing the Supersky flat to dawn flats, using the same methodology for analysis. 
As flat fields were only obtained in three bands, comparisons are shown for those bands. 
The differences between dawn flats on each night are within one percent, matching expectations. 
Similarly, Table~\ref{tab:sp2} compares the Supersky flat to dusk flats, also showing percent level agreement with the dusk reference each night. 
All flat fields show sub-percent level consistency. 
The Supersky flat field is used as an independent check of the  dusk and dawn flat field's ability to represent the illumination.
 \begin{table}[htbp]
    \caption{Supersky and different day flat comparison}  \label{tab:sp}
    \centering
    \begin{subtable}[t]{0.45\linewidth}
   	\centering
   	\subcaption{Supersky and different dawn flat comparison}
        \label{tab:sp1}
    \begin{tabular}{cccc}
		\hline\noalign{\smallskip}
		Date(YYYYMMDD)/Filter & B \% & V \% & R \% \\
		\hline\noalign{\smallskip}
			20220924 & 0.3 & 0.3 & 0.4\\
			20220925 & 0.7 & 0.8 & 0.8\\
			20220926 & 0.4 & 0.4 & 0.5\\
			20220927 & 0.4 & 0.3 & 0.4\\
			20220928 & 0.4 & 0.4 & 0.4\\
			20220929 & 0.4 & 0.4 & 0.4\\
		\noalign{\smallskip}\hline
        \end{tabular}
        \end{subtable}
    \hfill
    \begin{subtable}[t]{0.45\linewidth}
    \centering
    \subcaption{Supersky and different dusk flat comparison}
        \label{tab:sp2}
	\begin{tabular}{cccc}
	\hline\noalign{\smallskip}
		Date(YYYYMMDD)/Filter & B \% & V \% & R \%  \\
		\hline\noalign{\smallskip}
			20220924 & 0.4 & 0.6 & 0.7\\
			20220925 & 0.4 & 0.5 & 0.6\\
			20220926 & 0.5 & 0.5 & 0.5\\
			20220927 & 0.4 & 0.4 & 0.6\\
			20220928 & 0.3 & 0.3 & 0.3\\
			20220929 & 0.6 & 0.7 & 0.9\\
		 \noalign{\smallskip}\hline
        \end{tabular}
        \end{subtable}
\end{table}

Table~\ref{tab:sp} demonstrates that the flat field stable at slightly under 0.01 over nights. 
Furthermore, each band flat field and the Supersky flat field remain within 1\% over the entire observation period. This is in contrast to the Supersky flat field for the 60 cm observation field. 
It suggests that changes in observing conditions like airmass or twilight illumination are less than 0.01, validated by the independent Supersky flat. 

\subsection{Limiting Magnitude}
\begin{figure}[htbp]
  \centering
    \begin{subfigure}[t]{0.45\textwidth}
    \includegraphics[width=75mm]{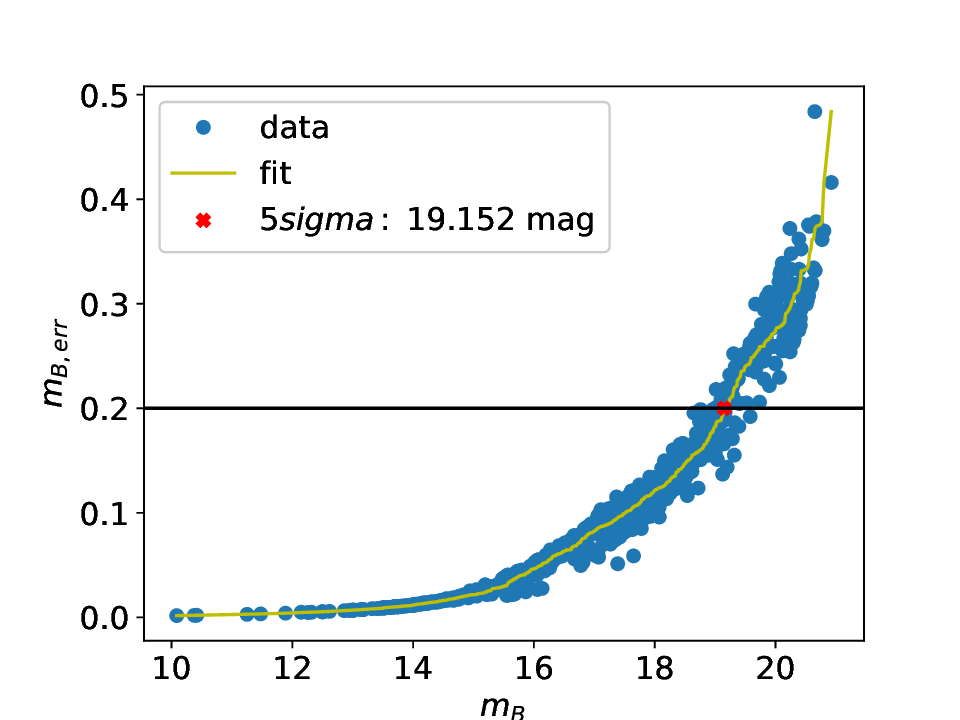}
    \caption{\label{fig:limmB}{\small B-band magnitude-error plot}}
  \end{subfigure}
  \hfill
  \begin{subfigure}[t]{0.45\linewidth}
  \centering
   \includegraphics[width=75mm]{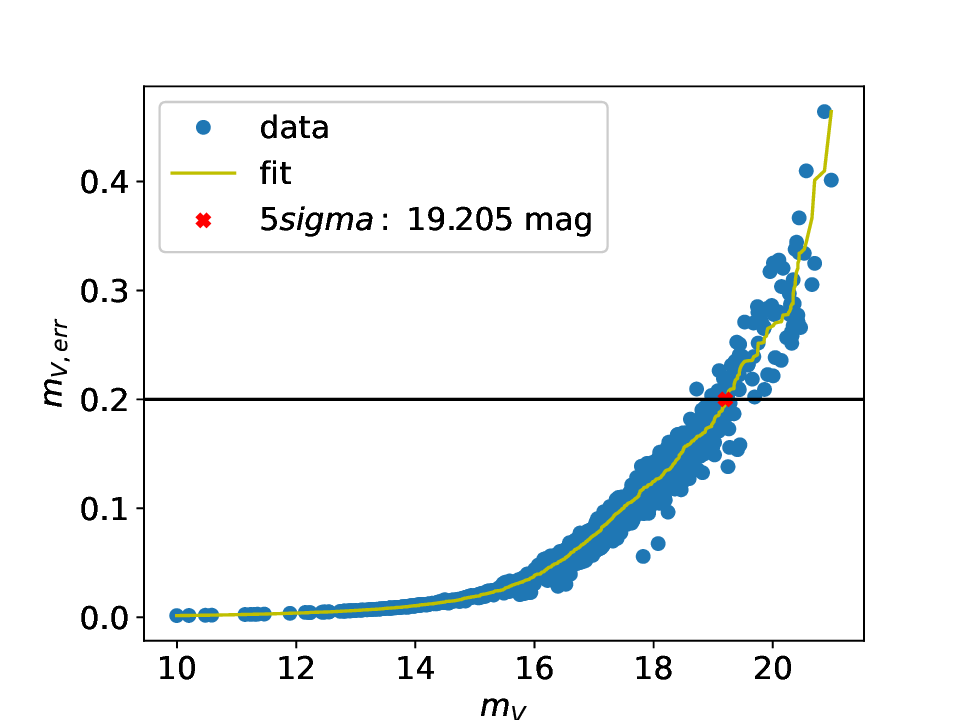}
	  \caption{\label{fig:limmV}{\small V-band magnitude-error plot} }
  \end{subfigure}
  \vfill
  \begin{subfigure}[t]{0.45\textwidth}
  \centering
   \includegraphics[width=75mm]{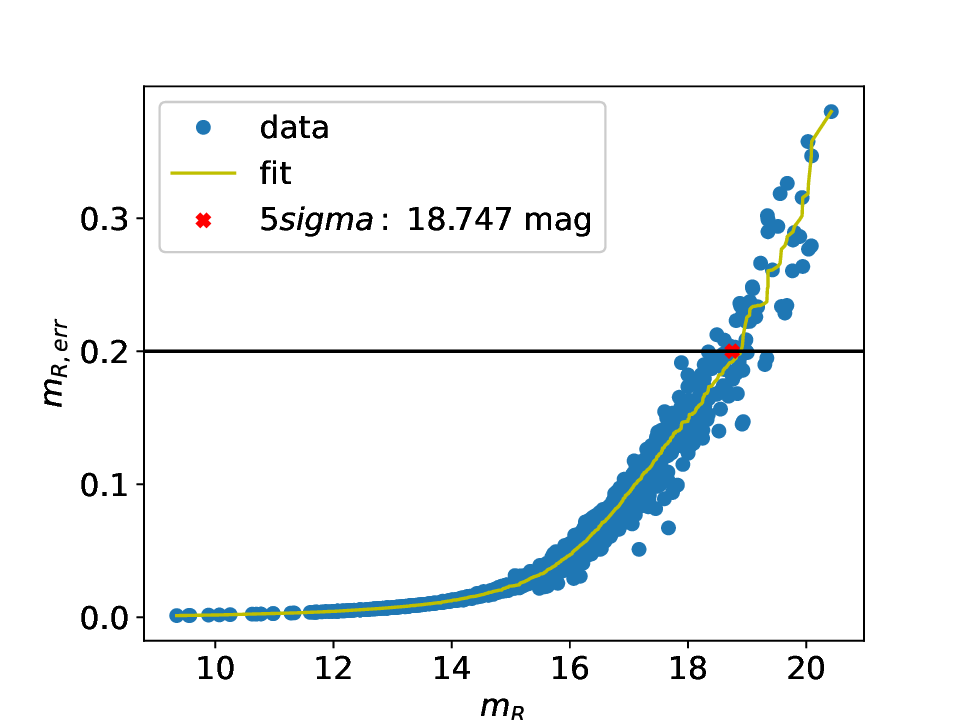}
	  \caption{\label{fig:limmR}{\small R-band magnitude-error plot}}
    \end{subfigure}
    \hfill
  \begin{subfigure}[t]{0.45\textwidth}
    \includegraphics[width=75mm]{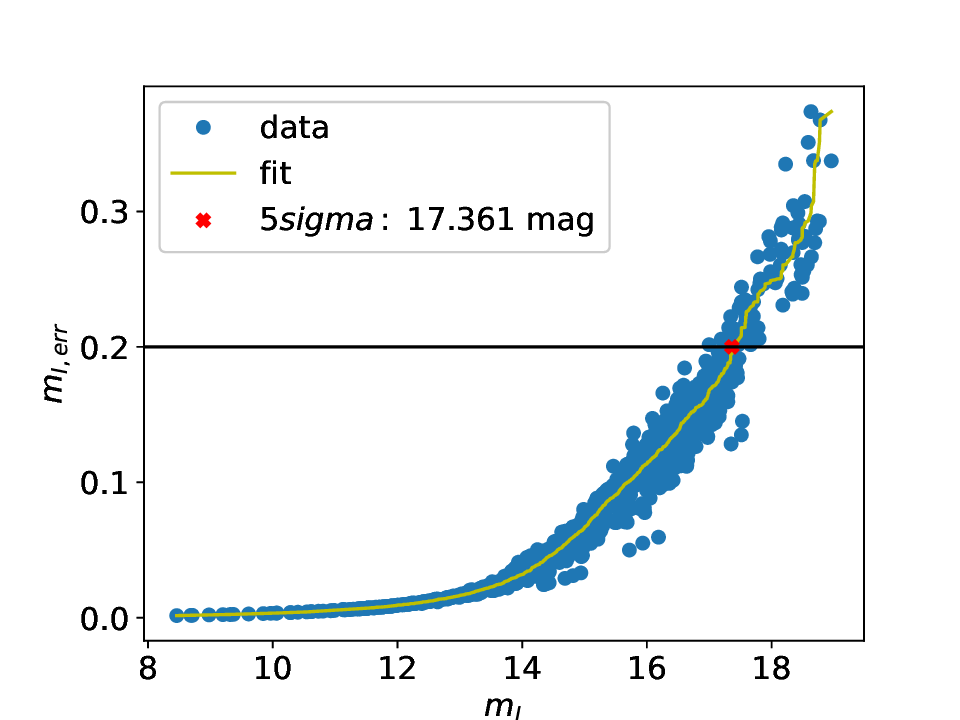}
    \caption{\label{fig:limmI}{\small I-band magnitude-error plot}}
  \end{subfigure}
  \caption{The $5\sigma$ SNR limiting magnitude of each band at 30 second exposures.The calibrated magnitude on x axis and magnitude error on y axis. The blue data points represent the observed data, the yellow curve is a fitted curve to the magnitude error, and the red markers are points with an SNR of $5\sigma$. The plot shows that the limiting magnitude is 19.152 $mag$ in B-band,  19.205 $mag$ in V-band, 18.747 $mag$ in R-band and 17.361 $mag$ in I-band.}
  \label{fig:mag_err}
\end{figure}

\hspace{2em}The limiting magnitude is an important indicator of an observing system, showing what the observing system's capability. In this system, due to the mechanical limitations of the telescope, we can only use a single exposure up to 30 seconds. 
Below are the limiting magnitudes calculated by the open clusters, after calibrating, and calculating the limiting magnitudes in the $BVRI$ bands and at $3\sigma $ and $5 \sigma$. 
Figure \ref{fig:mag_err} shows the magnitude-error plots and magnitude statistics after the calibration. Here the $m_{err}$ is obtained from Equation \ref{eq:fluxerr}.\\
\begin{table}[htbp]
	\begin{center}
		\caption{Limiting magnitudes at $3\sigma $ and $5 \sigma$ for each band}		\label{lm}
		\begin{tabular}{cccc}
		\hline\noalign{\smallskip}
		{Filter} & {$5 \sigma (mag)$} & Accuracy of Fitting & {Exposure \ time(s)}  \\
		\hline\noalign{\smallskip}
			B & 19.152 & 0.042 & 30\\
			V & 19.205 & 0.043 & 30\\
			R & 18.747 & 0.053 & 30\\
			I & 17.361 & 0.052 & 30\\
			\noalign{\smallskip}\hline
		\end{tabular}
	\end{center}
\end{table}
From the table\ref{lm} we see that limiting magnitude variation in different bands corresponds to the quantum efficiency of the CMOS, and the limiting magnitude at $5 \sigma$ in the V band is $19.205 \ mag$, as our expectations.

\subsection{Open Cluster}
\begin{figure}[htbp]
  \centering
    \begin{subfigure}[t]{0.45\textwidth}
    \includegraphics[width=75mm]{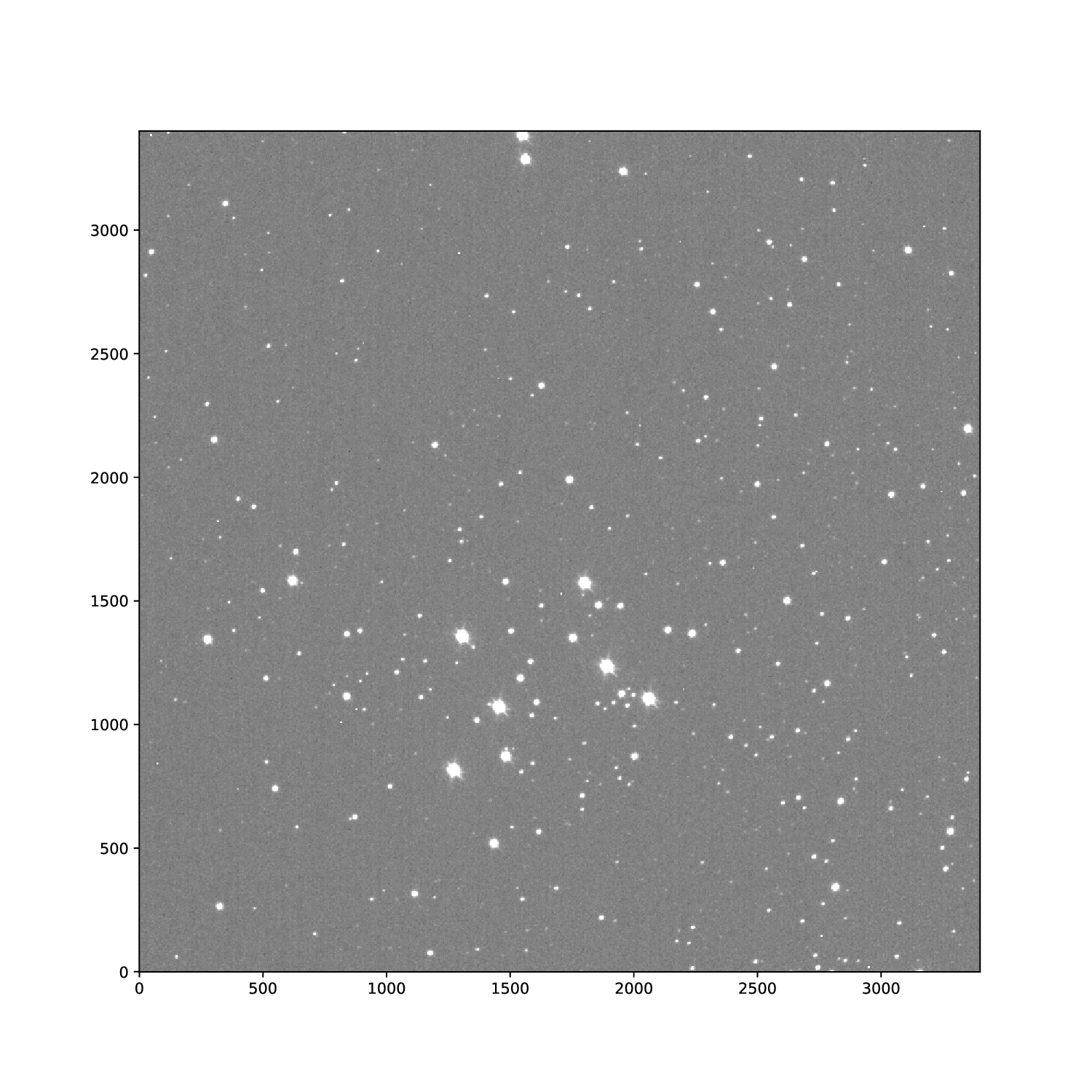}
    \caption{\label{fig:ngc6913-ri}{\small NGC6913 R-band}}
  \end{subfigure}
  \hfill
  \begin{subfigure}[t]{0.45\linewidth}
  \centering
   \includegraphics[width=75mm]{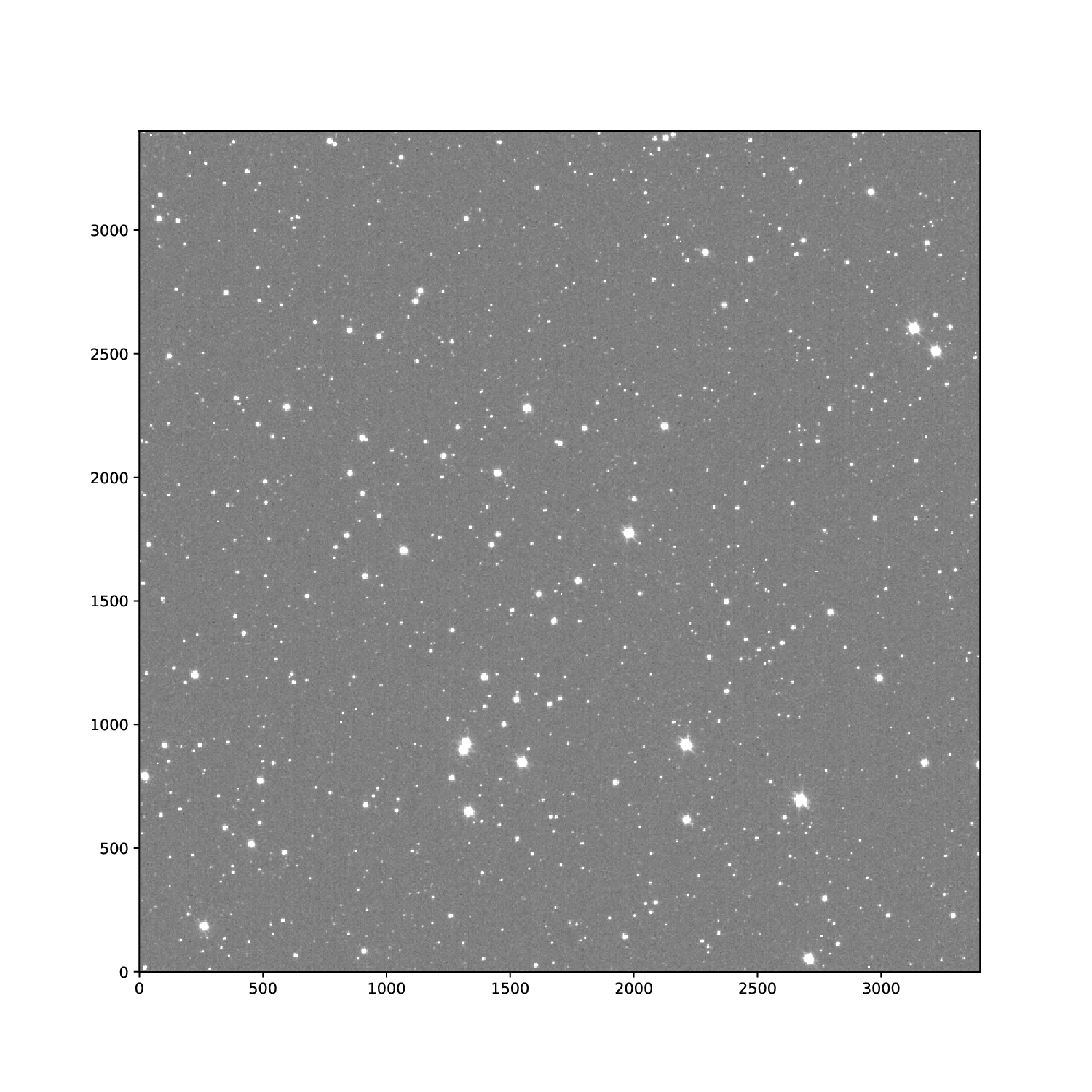}
	  \caption{\label{fig:ngc7243-ri}{\small NGC7243 R-band} }
  \end{subfigure}%
  \vfill
  \begin{subfigure}[t]{0.45\textwidth}
  \centering
   \includegraphics[width=75mm]{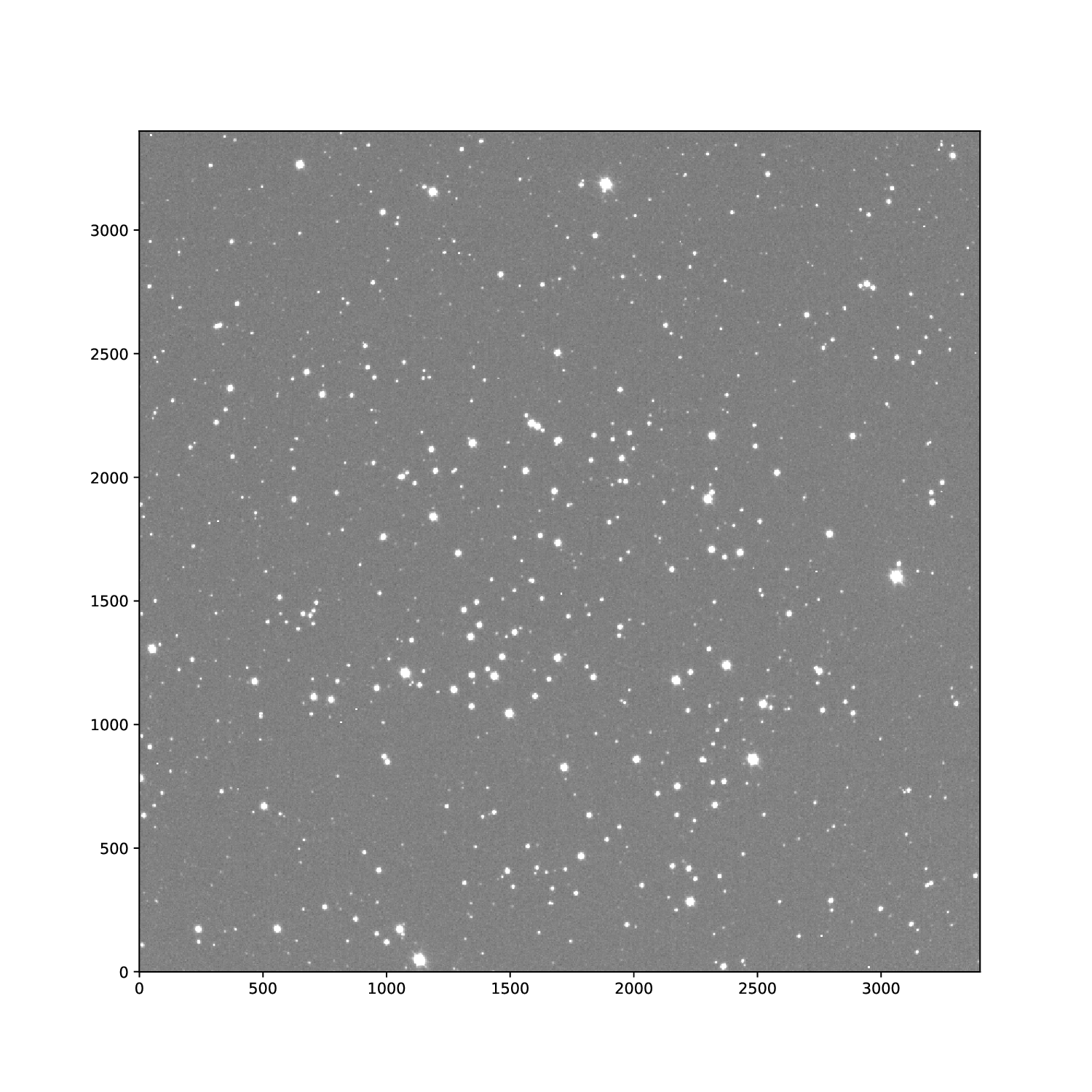}
	  \caption{\label{fig:ngc6811-ri}{\small NGC6811 R-band}}
    \end{subfigure}
     \hfill
  \begin{subfigure}[t]{0.45\linewidth}
  \centering
   \includegraphics[width=75mm]{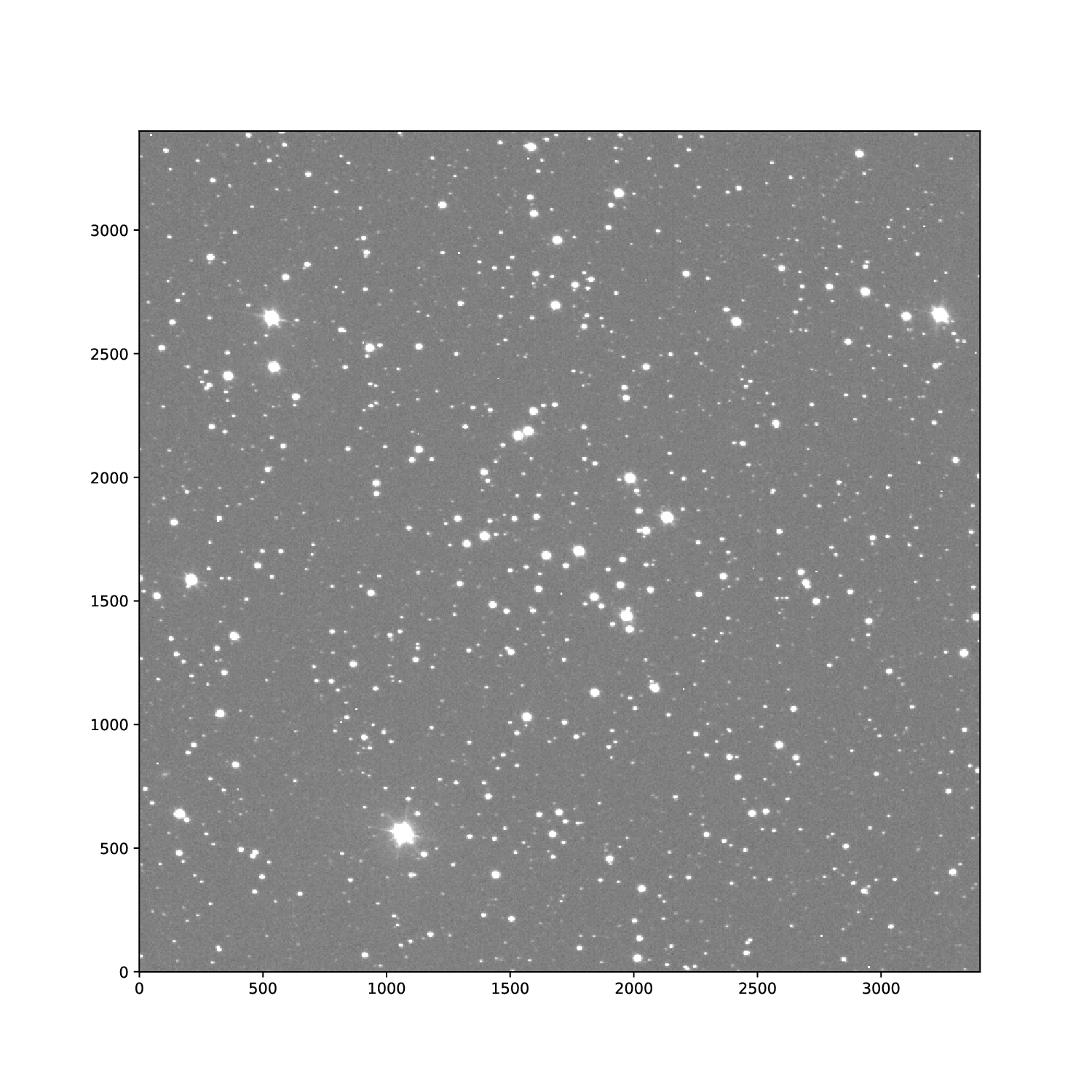}
	  \caption{\label{fig:ngc744-ri}{\small NGC744 R-band} }
  \end{subfigure}%
  \caption{R-band images of each open clusters}
  \label{fig:trimming-ngc-r}
\end{figure}

\hspace{2em}Table \ref{xw} shows four open star clusters that we observed, namely NGC6913, NGC7243, NGC6811 and NGC744. Figure \ref{fig:trimming-ngc-r} shows their R-band trimming images. In this section, we will separately introduce the imaging and processing results of each open star cluster.

After completing the photometry, the reference magnitude we used was Gaia's synthetic photometry. Based on the characteristics of open clusters, since in the same field, the extinction between the individual stars can be almost equal, so we regard $k_1 \times X$ as a constant, and thus the Equation \ref{fum:mian} becomes:

	\begin{equation}
	\begin{aligned}
	     m_{B,inst}=m_{B,Gaia-sp}+c_{B}+constant + k_{2,B} \cdot (m_{B,Gaia-sp}-m_{V,Gaia-sp})\\
		m_{V,inst}=m_{V,Gaia-sp}+c_{V}+constant + k_{2,V} \cdot (m_{B,Gaia-sp}-m_{V,Gaia-sp})\\
		m_{R,inst}=m_{R,Gaia-sp}+c_{R}+constant + k_{2,R} \cdot (m_{V,Gaia-sp}-m_{R,Gaia-sp})\\
		m_{I,inst}=m_{I,Gaia-sp}+c_{I}+constant + k_{2,I} \cdot (m_{V,Gaia-sp}-m_{I,Gaia-sp})
		\end{aligned}
		\label{eq:oc}
	\end{equation}
We can see that Equation \ref{eq:oc} have only two variations, $C + constant$ and $k_2$, from which we can obtain their respective corresponding parameters by simple straight line fitting. 
The color correction are combined using the Equation \ref{eq:oc}(\citealt{Cousins+etal+2001};\citealt{Calamida+etal+2018}).
\\\\
\textit{NGC6913}

\begin{figure}[htbp]
  \centering
    \begin{subfigure}[t]{0.45\textwidth}
    \includegraphics[width=75mm]{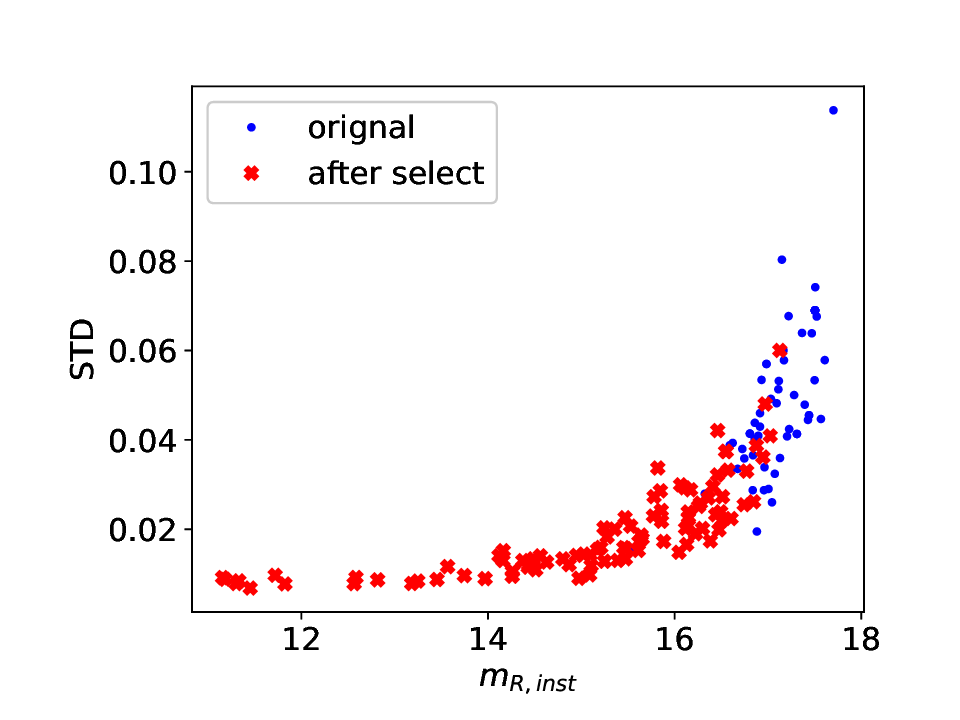}
    \caption{\label{fig:NGC6913-I-r}{\small }}
  \end{subfigure}
  \hfill
  \begin{subfigure}[t]{0.45\linewidth}
  \centering
   \includegraphics[width=75mm]{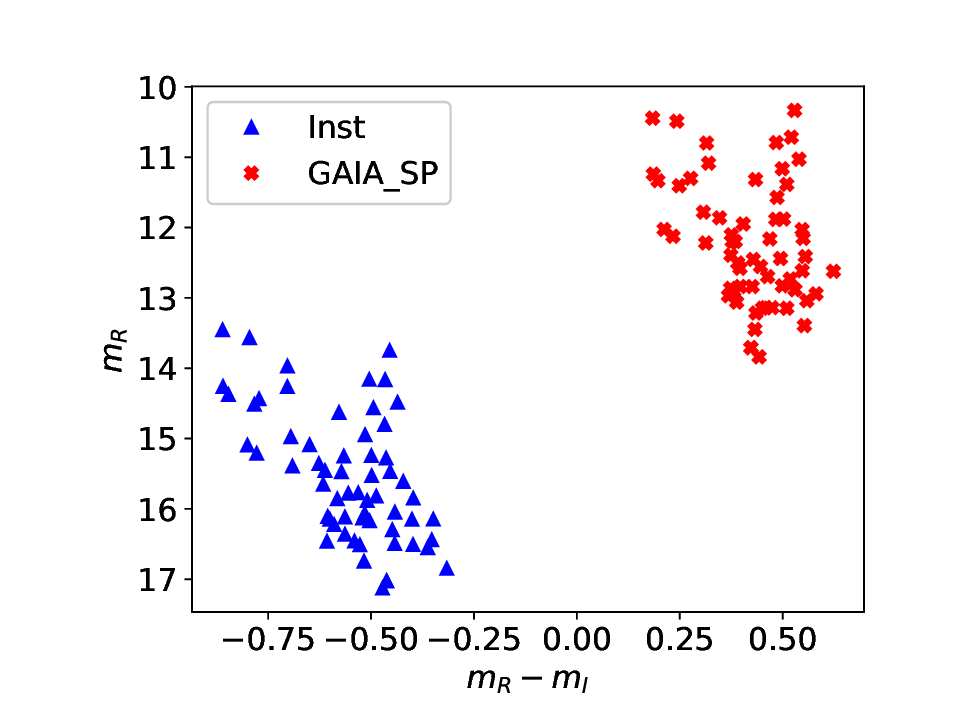}
	  \caption{\label{fig:NGC6913-B-r}{\small } }
  \end{subfigure}%
  \vfill
  \begin{subfigure}[t]{0.45\textwidth}
  \centering
   \includegraphics[width=75mm]{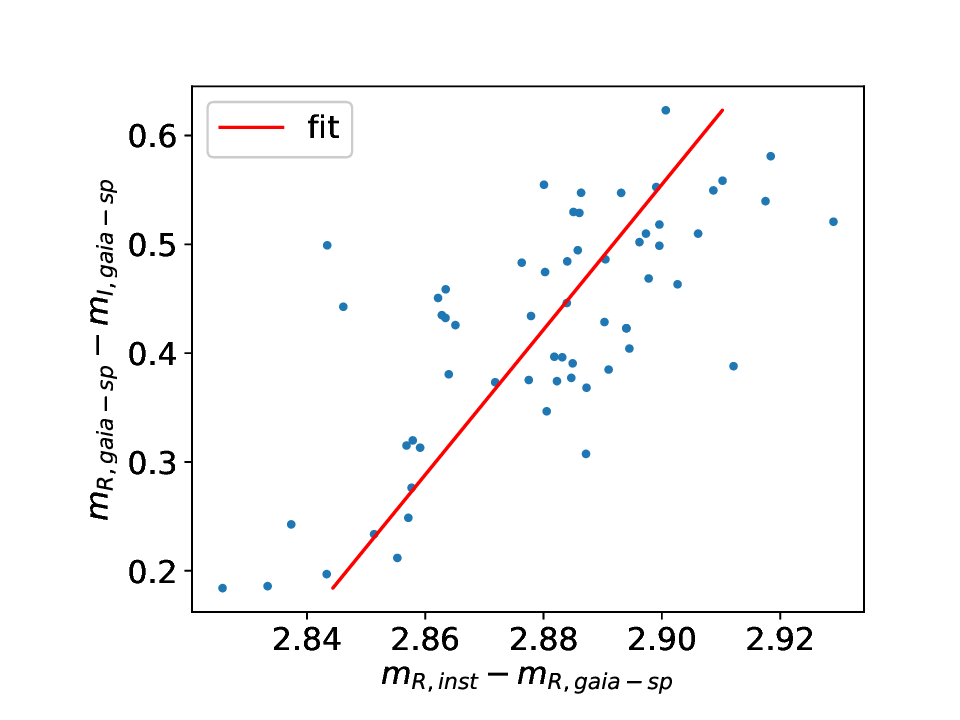}
	  \caption{\label{fig:NGC6913-V-r}{\small }}
    \end{subfigure}
    \hfill
  \begin{subfigure}[t]{0.45\textwidth}
    \includegraphics[width=75mm]{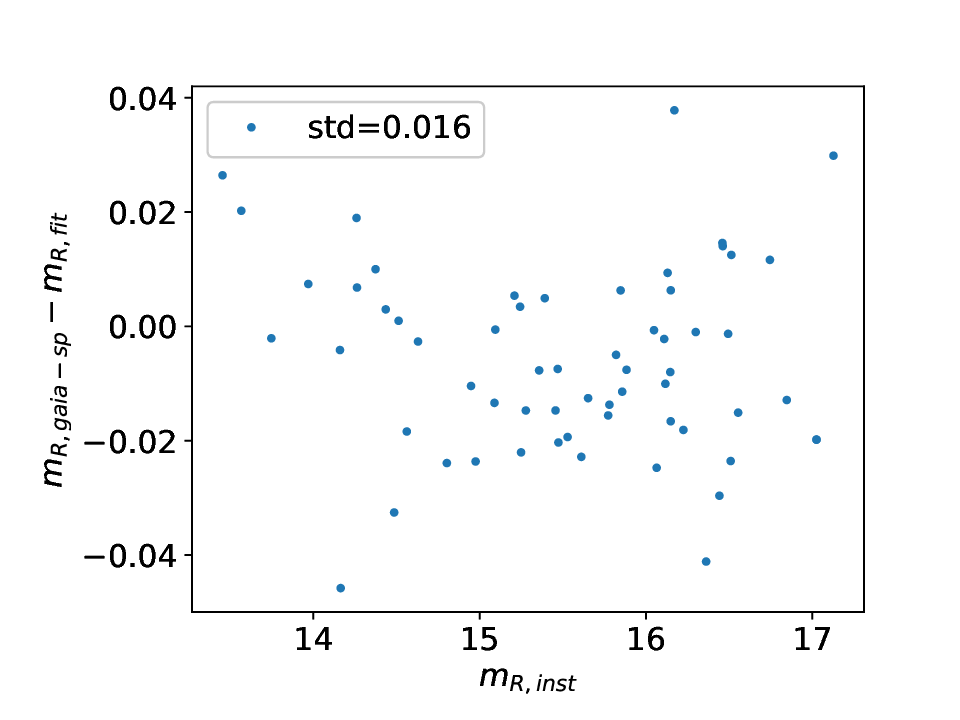}
    \caption{\label{fig:NGC6913-R-r}{\small }}
  \end{subfigure}
  \caption{NGC6913 R-band results,(a): The NGC6913 R-band's magnitude-STD diagram, with red crosses representing the selected data points and blue dots representing the excluded data points. (b): The NGC6913 R-band's color-magnitude diagram, where the blue triangle represents the instrumental magnitude and the red square corresponds to the Gaia-SP magnitude. The horizontal axis is $m_R-m_I$ and the vertical axis is $m_R$. (c): The NGC6913 R-band's magnitude-color diagram, with blue dots representing data points and a red line representing the fitted line. The horizontal axis of the plot represents the difference between instrumental magnitude and Gaia-SP magnitude, while the vertical axis represents Gaia-SP's color, denoted as $m_{R,GAIA-SP}-m_{I,GAIA-SP}$, as described in equation \ref{eq:oc}. (d):The NGC6913 R-band's magnitude-error diagram, with blue dots representing data points. The horizontal axis represents instrumental magnitude, while the vertical axis represents the difference between instrumental magnitude and the fit magnitude according to Equation \ref{eq:oc}.}
  \label{fig:ngc6913-r}
\end{figure}

We performed photometry on 80 frame in the same band, calculated the standard deviation of the magnitude distribution for each star over these 80 frame, and shown  Figure \ref{fig:NGC6913-I-r}. 
It displays the error of stars before our selection, with instrumental magnitude on the x-axis and the standard deviation of the magnitude distribution for each source across the 80 frame on the y-axis. 
Based on Figure \ref{fig:NGC6913-I-r}, we obtained an internal precision of magnitude measurement. 
We then used Figure \ref{fig:NGC6913-I-r} for target selection, removing the variable stars, saturation stars, and  Non Zero Age Main-Sequence. 
Finally, we selected member stars and matched the final selected sources to the GAIA-SP star catalog, showing in Figure \ref{fig:NGC6913-I-r}. 

Figure \ref{fig:NGC6913-B-r} compares the color-magnitude plot of the instrumental magnitude with the color-magnitude plot of Gaia-SP, the horizontal axis is the R-I color, and the vertical axis is the R-band magnitude; Figure \ref{fig:NGC6913-V-r} is from Equation\ref{eq:oc}.

And Figure \ref{fig:NGC6913-R-r} the vertical axis is difference between the corrected magnitude $m_{R,fit}$ and Gaia-SP $m_{Gaia-sp}$. 
After the final fit, we calculate the difference between the magnitude and Gaia-SP color by multiplying the color coefficient with the color value. 
The R-band instrumental magnitude is the horizontal axis, and the accuracy obtained from the final photometry is reflected in the standard deviation(std) of the label in the figure \ref{fig:NGC6913-R-r}. 
Table \ref{tab:ngc_R} displays the coefficients $k_{2,R}$, zeroes $c_{R,0}+constant$, and photometric accuracy that were computed for each open cluster. From this table, we can see that its photometric accuracy is 0.016mag. \\
\\
\textit{NGC7243}

\begin{figure}[htbp]
  \centering
    \begin{subfigure}[t]{0.45\textwidth}
    \includegraphics[width=75mm]{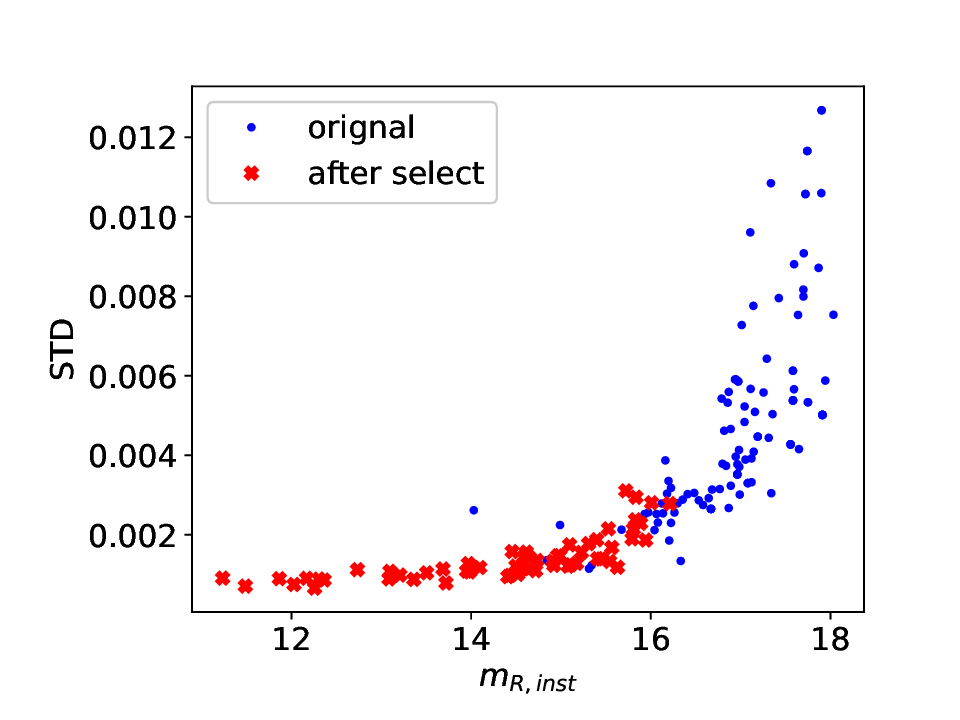}
    \caption{\label{fig:NGC7243-I}}
  \end{subfigure}
  \hfill
  \begin{subfigure}[t]{0.45\linewidth}
  \centering
   \includegraphics[width=75mm]{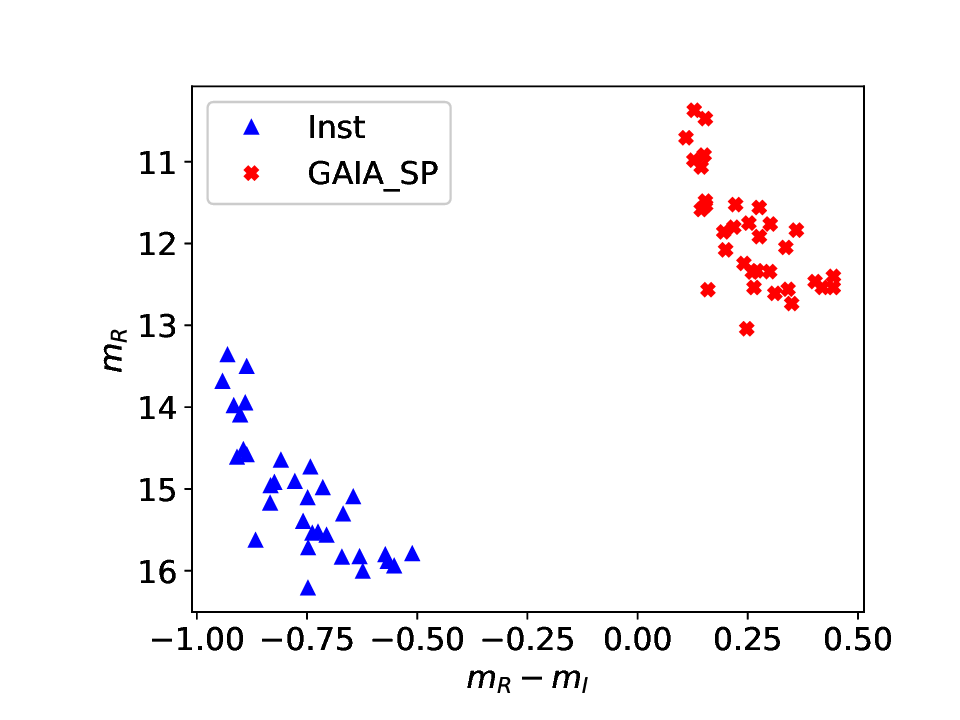}
	  \caption{\label{fig:NGC7243-B}}
  \end{subfigure}
  \vfill
  \begin{subfigure}[t]{0.45\textwidth}
  \centering
   \includegraphics[width=75mm]{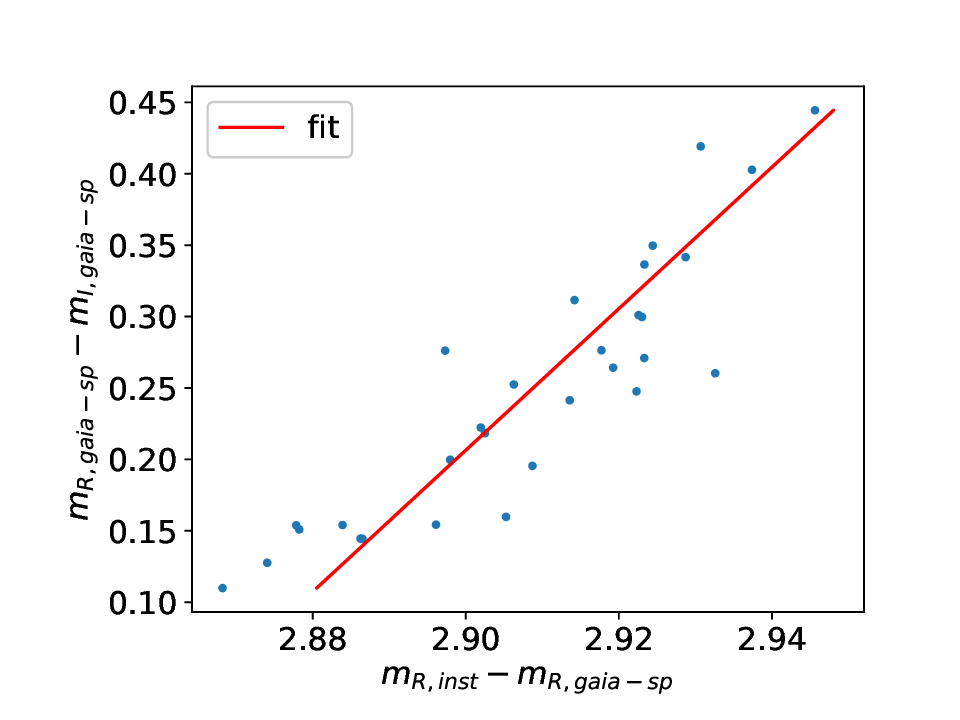}
	  \caption{\label{fig:NGC7243-V}}
    \end{subfigure}
    \hfill
  \begin{subfigure}[t]{0.45\textwidth}
    \includegraphics[width=75mm]{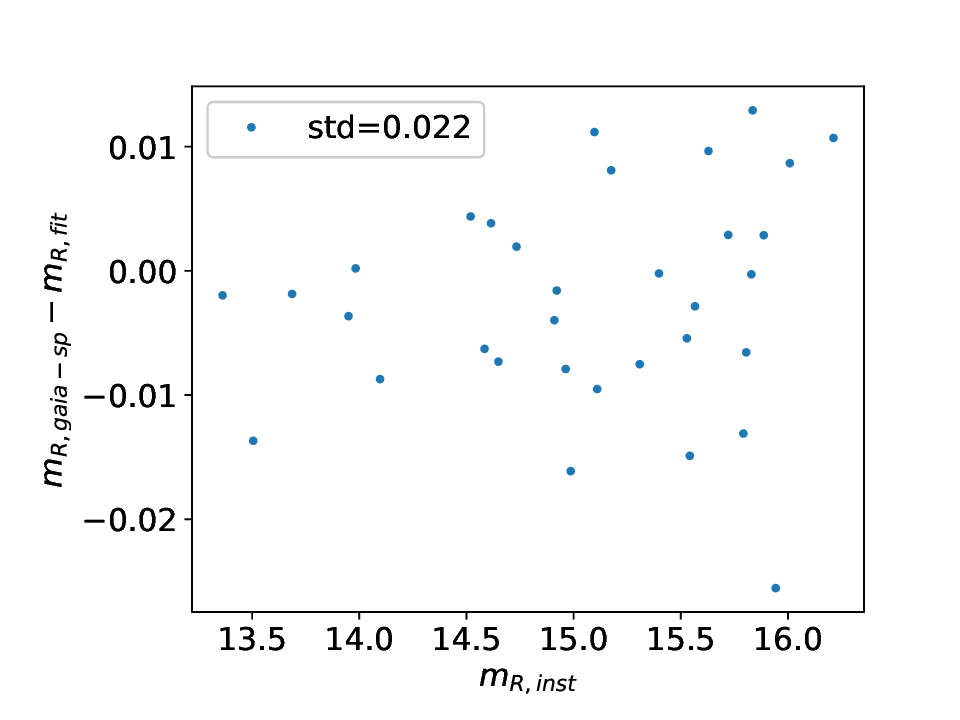}
    \caption{\label{fig:NGC7243-R}}
  \end{subfigure}
  \caption{Some as in Figure \ref{fig:ngc6913-r} NGC7243 R-band results, (a): The R-band's magnitude-STD diagram, with red crosses representing the selected data points and blue dots representing the excluded data points. (b): The R-band's color-magnitude diagram, where the blue triangle represents the instrumental magnitude and the red square is the Gaia-SP magnitude. The horizontal axis is $m_R-m_I$ and the vertical axis is $m_R$. (c): The R-band's magnitude-color diagram, with blue dots representing data points and a red line representing the fitted line. The horizontal axis of the plot represents the difference between instrumental magnitude and Gaia-SP magnitude, while the vertical axis represents Gaia-SP's color, denoted as $m_{R,GAIA-SP}-m_{I,GAIA-SP}$, as described in equation \ref{eq:oc}. (d):The R-band's magnitude-error diagram, with blue dots representing data points. The horizontal axis represents instrumental magnitude, while the vertical axis represents the difference between instrumental magnitude and the fitted magnitude according to Equation \ref{eq:oc}.}
  \label{fig:ngc7243}
\end{figure}

The processing procedure for NGC7243 is similar to that of NGC6913.
Figure \ref{fig:NGC7243-I} shows the magnitude and standard deviation plot obtained with the same processing procedure for NGC6913. 
Finally, by calculating the standard deviation of the data in the scatter plot of Figure  \ref{fig:NGC7243-V}, we obtained an photometric accuracy of 0.022 $mag$.
\\\\
\textit{NGC6811}

\begin{figure}[htbp]
  \centering
    \begin{subfigure}[t]{0.45\textwidth}
    \includegraphics[width=75mm]{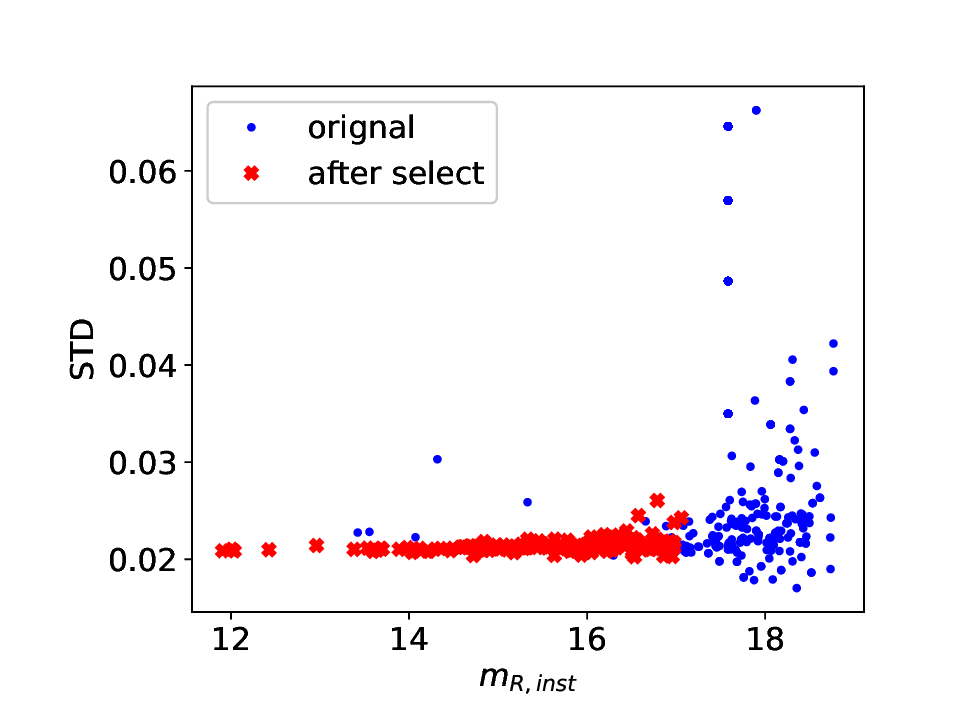}
    \caption{\label{fig:NGC6811-I}{\small }}
  \end{subfigure}
  \hfill
  \begin{subfigure}[t]{0.45\linewidth}
  \centering
   \includegraphics[width=75mm]{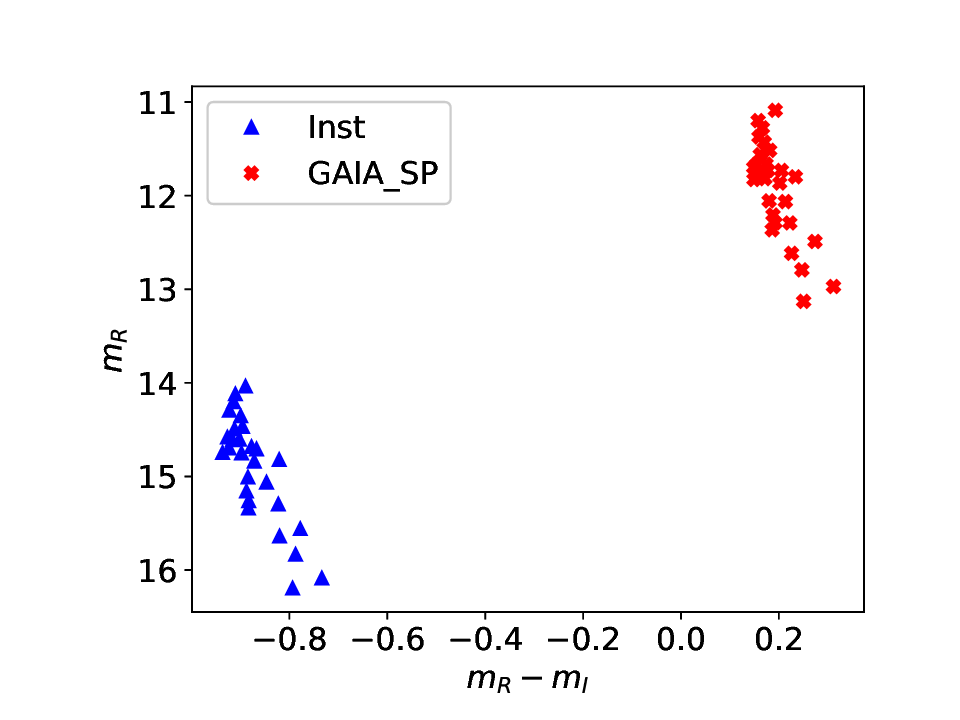}
	  \caption{\label{fig:NGC6811-B}{\small } }
  \end{subfigure}
  \vfill
  \begin{subfigure}[t]{0.45\textwidth}
  \centering
   \includegraphics[width=75mm]{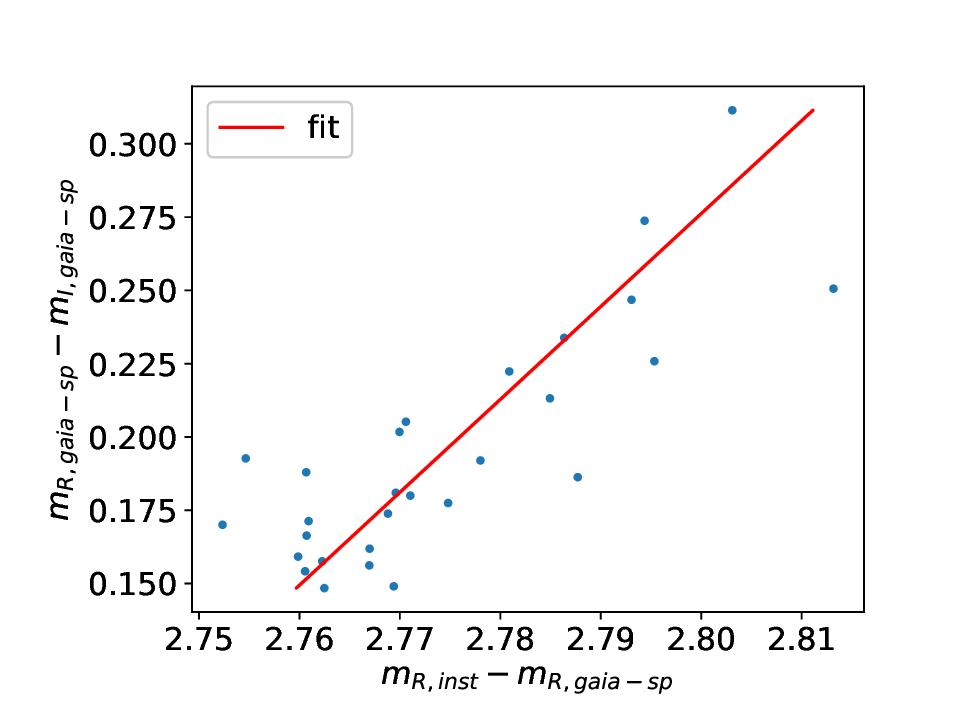}
	  \caption{\label{fig:NGC6811-V}{\small }}
    \end{subfigure}
    \hfill
  \begin{subfigure}[t]{0.45\textwidth}
    \includegraphics[width=75mm]{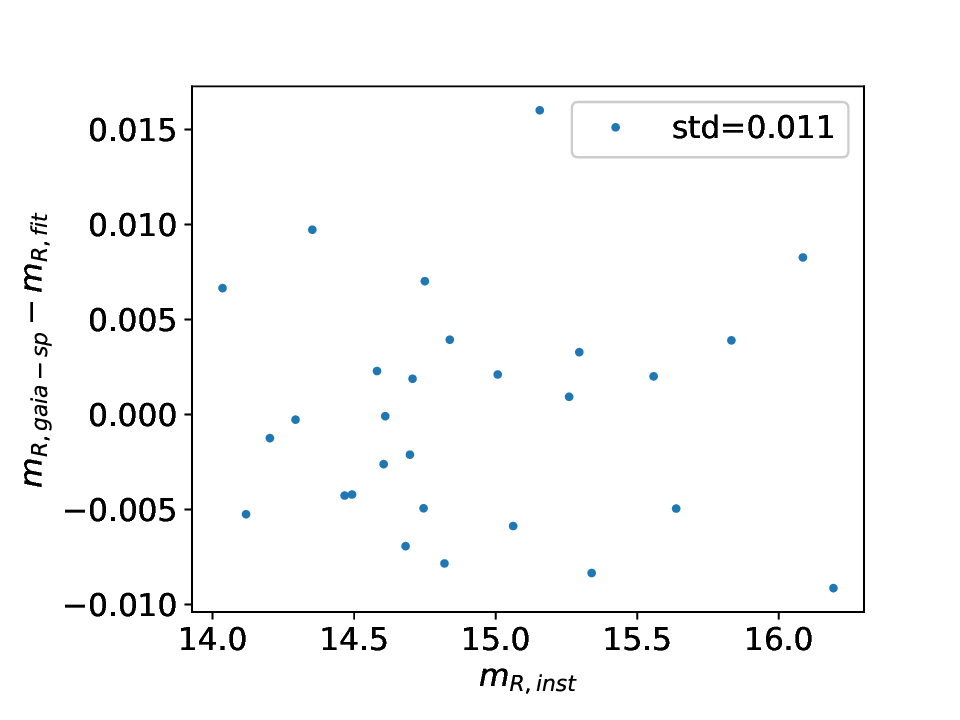}
    \caption{\label{fig:NGC6811-R}{\small }}
  \end{subfigure}
  \caption{Some as in Figure \ref{fig:ngc6913-r} NGC6811 R-band results.(a): The R-band's magnitude-STD diagram, with red crosses representing the selected data points and blue dots representing the excluded data points. (b): The R-band's color-magnitude diagram, where the blue triangle represents the instrumental magnitude and the red square is the Gaia-SP magnitude. The horizontal axis is $m_R-m_I$ and the vertical axis is $m_R$. (c): The R-band's magnitude-color diagram, with blue dots representing data points and a red line representing the fitted line. The horizontal axis of the plot represents the difference between instrumental magnitude and Gaia-SP magnitude, while the vertical axis represents Gaia-SP's color, denoted as $m_{R,GAIA-SP}-m_{I,GAIA-SP}$, as described in equation \ref{eq:oc}. (d):The R-band's magnitude-error diagram, with blue dots representing data points. The horizontal axis represents instrumental magnitude, while the vertical axis represents the difference between instrumental magnitude and the fitted magnitude according to Equation \ref{eq:oc}.}
  \label{fig:ngc6811}
\end{figure}

The processing procedure for NGC6811 is similar to that of NGC6913.
Figure \ref{fig:NGC6811-I} shows the magnitude and standard as same processing procedure of NGC6913. Figure \ref{fig:ngc6811} also represents the same calculation process as Figure \ref{fig:ngc6811} of NGC6913. 
Finally, by calculating the standard deviation of the data in the scatter plot of Figure \ref{fig:NGC6811-V}, with photometric accuracy of 0.011 $mag$.
\\\\
\textit{NGC744}

\begin{figure}[htbp]
  \centering
    \begin{subfigure}[t]{0.45\textwidth}
    \includegraphics[width=75mm]{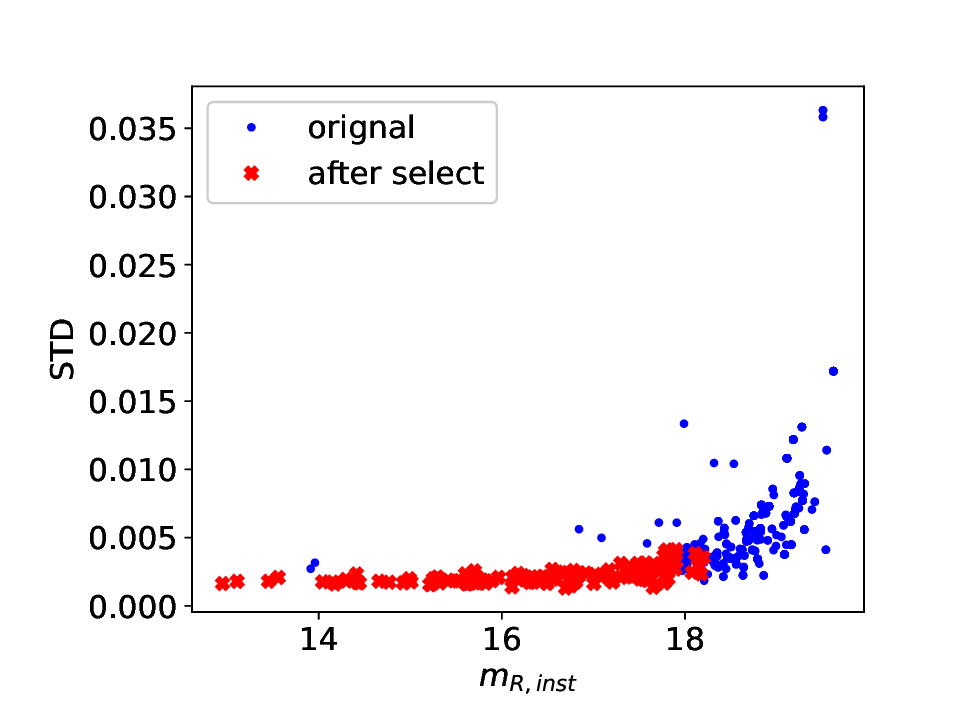}
    \caption{\label{fig:NGC744-I}{\small }}
  \end{subfigure}
  \hfill
  \begin{subfigure}[t]{0.45\linewidth}
  \centering
   \includegraphics[width=75mm]{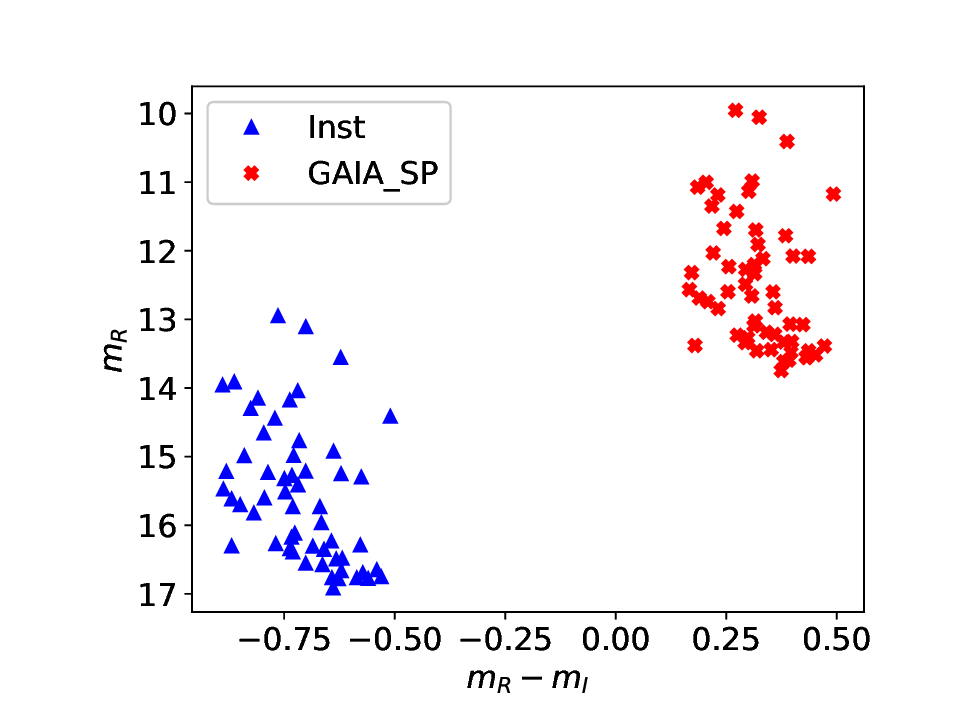}
	  \caption{\label{fig:NGC744-B}{\small } }
  \end{subfigure}
  \vfill
  \begin{subfigure}[t]{0.45\textwidth}
  \centering
   \includegraphics[width=75mm]{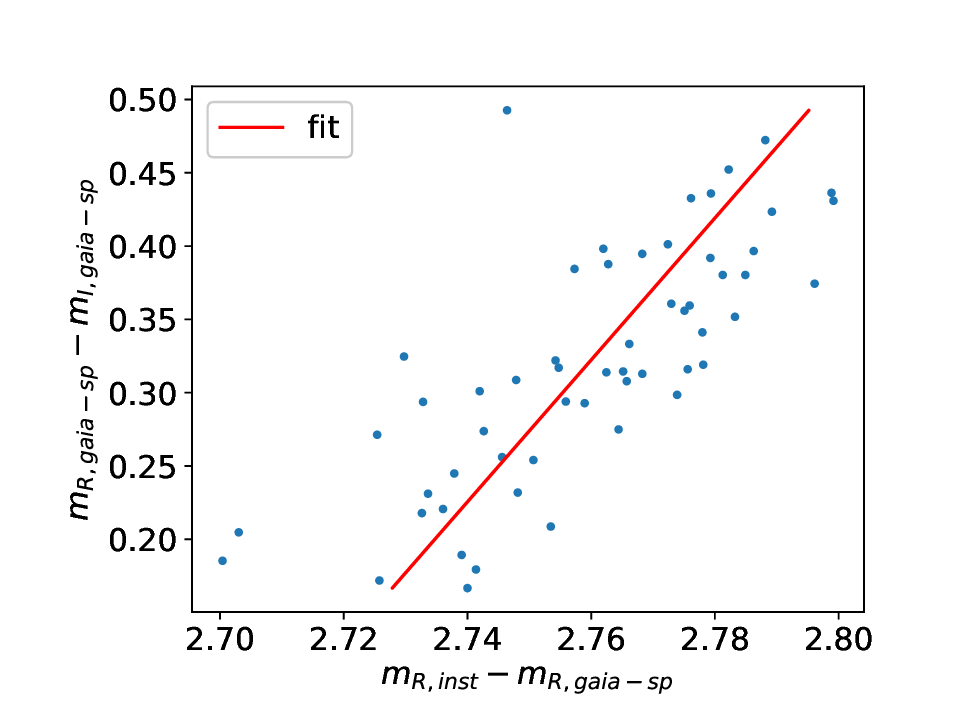}
	  \caption{\label{fig:NGC744-V}{\small }}
    \end{subfigure}
    \hfill
  \begin{subfigure}[t]{0.45\textwidth}
    \includegraphics[width=75mm]{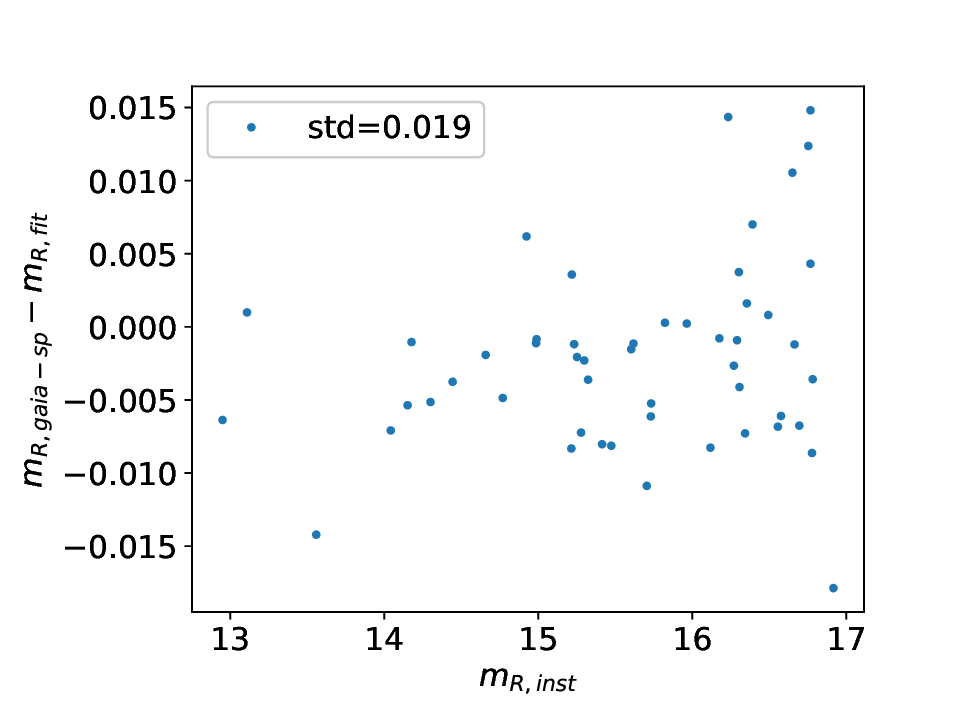}
    \caption{\label{fig:NGC744-R}{\small }}
  \end{subfigure}
  \caption{Some as in Figure \ref{fig:ngc6913-r} NGC744 R-band results.(a): The R-band's magnitude-STD diagram, with red crosses representing the selected data points and blue dots representing the unselected data points. (b): The R-band's color-magnitude diagram, where the blue triangle represents the instrumental magnitude and the red square is the Gaia-SP magnitude. The horizontal axis is $m_R-m_I$ and the vertical axis is $m_R$. (c): The R-band's magnitude-color diagram, with blue dots representing data points and a red line representing the fitted line. The horizontal axis of the plot represents the difference between instrumental magnitude and Gaia-SP magnitude, while the vertical axis represents Gaia-SP's color, denoted as $m_{R,GAIA-SP}-m_{I,GAIA-SP}$, as described in equation \ref{eq:oc}. (d):The R-band's magnitude-error diagram, with blue dots representing data points. The horizontal axis represents instrumental magnitude, while the vertical axis represents the difference between instrumental magnitude and the magnitude obtained by fitting according to Equation \ref{eq:oc}.}
  \label{fig:ngc744r}
\end{figure}

The processing procedure for NGC744 is similar to that of NGC6913. Figure \ref{fig:NGC744-I} shows the magnitude and standard deviation plot obtained through the same processing procedure as NGC6913. Figure \ref{fig:ngc744r} also represents the same calculation process as Figure \ref{fig:ngc744r} of NGC744. Finally, by calculating the standard deviation of the data in the scatter plot of Figure \ref{fig:NGC744-R}, we obtained an photometric accuracy of 0.019 $mag$.

Table \ref{tab:ngc_R} summarizes the results for these four open star clusters, where the standard deviation listed in the table represents the photometric accuracy. 
The color coefficient $k_{2}$ is not the true system conversion coefficient and incorporates the main order contribution of the cluster Therefore, the photometric accuracy is approximately 0.02 $mag$(\citealt{Munari+etal+2014,Landolt+2007,Huang+etal+2021}).
\begin{table}[htbp]
	\begin{center}
		\caption{the result about the observed open cluster,$k_{2.R}$ and $c_{0,R}+constant$ correspond to the parameters of Equation \ref{eq:oc}.}		\label{tab:ngc_R}
		\begin{tabular}{cccc}
		\hline\noalign{\smallskip}
		{ID} & {$k_{2,R}$} & {$c_{R,0}+constant$} & {$\sigma$} \\
		\hline\noalign{\smallskip}
			NGC6913 & 0.208 & 2.789 & 0.016\\
			NGC7243 & 0.168 & 2.864 & 0.022\\
			NGC6811 & 0.149 &  2.734 & 0.011\\
			NGC744 & 0.208 & 2.682 & 0.019\\
			\noalign{\smallskip}\hline
		\end{tabular}
	\end{center}
\end{table}

\subsection{Standard Stars}
Table \ref{xw} shows that we observed two standard stars, namely SA 20-43 and HD 165434, in four different bands. 
After completing the pre-processing and photometric processing, we need to extract the light curves of the targets in each band, and then match them with the  catalog. 
After completing these steps, we calculate the extinction coefficients using the system transformation Equation\ref{eq:time}. 
At this point, the system transformation formula needs to be modified.Because there is only one target whose color is the same, for this standard star, the color term $k_2 \times (m_{B,Gaia}-m_{V,Gaia})$ can be regarded as a constant, and then the formula can be rewritten as:
\begin{equation}
	\begin{aligned}
	    m_{B,inst}=m_{B,Gaia-sp}+c_{B}+k_{1,B}\cdot X + constant\\
		m_{V,inst}=m_{V,Gaia-sp}+c_{V}+k_{1,V}\cdot X + constant \\
		m_{R,inst}=m_{R,Gaia-sp}+c_{R}+k_{1,R}\cdot X + constant\\
		m_{I,inst}=m_{I,Gaia-sp}+c_{I}+k_{1,I}\cdot X + constant
		\end{aligned}
		\label{eq:st}
\end{equation}

We can see that Equation\ref{eq:st} have only two variations, $c + constant$ and $k_1$, from which we obtain their respective corresponding parameters by linear fitting.
\\\\
\textit{HD 165434}

\begin{figure}[htbp]
  \centering
  \begin{subfigure}[t]{0.45\linewidth}
  \centering
   \includegraphics[width=75mm]{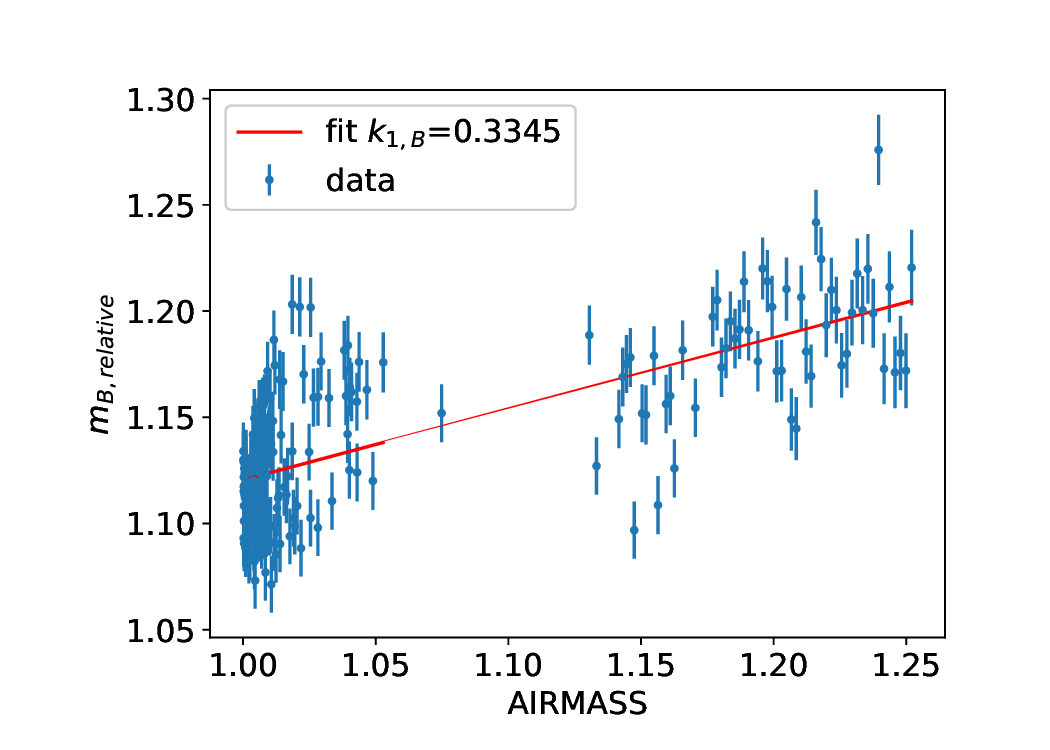}
	  \caption{\label{fig:hd165434-B}{\small B band} }
  \end{subfigure}%
  \hfill
  \begin{subfigure}[t]{0.45\textwidth}
  \centering
   \includegraphics[width=75mm]{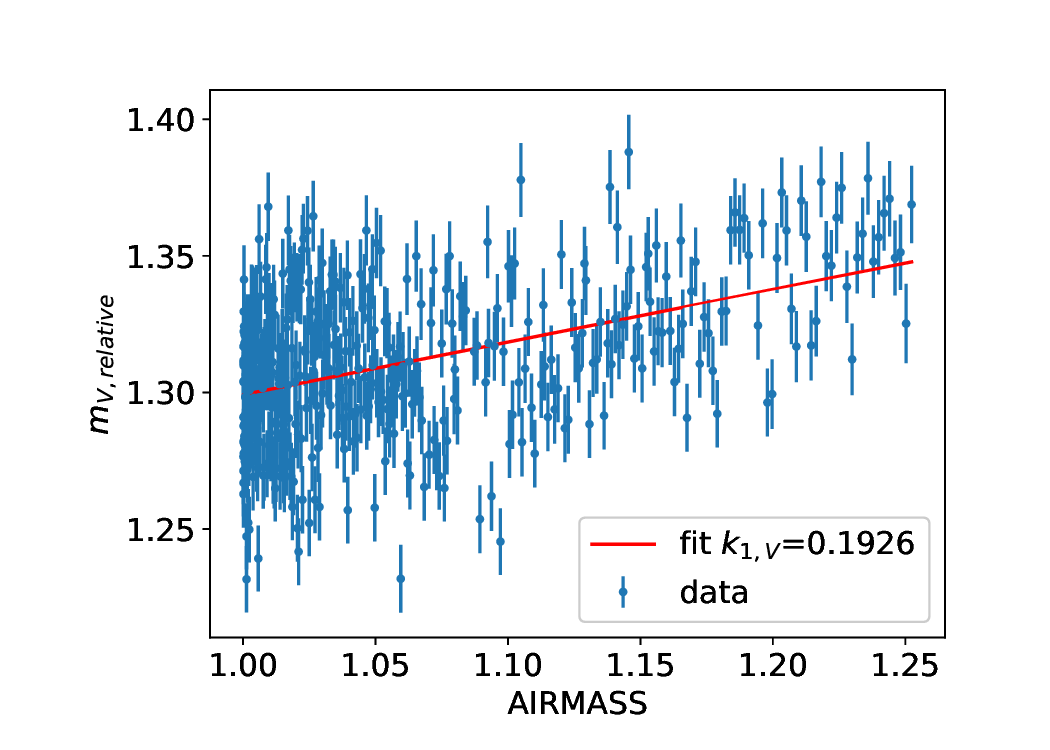}
	  \caption{\label{fig:hd165434-V}{\small V band}}
    \end{subfigure}
    \vfill
  \begin{subfigure}[t]{0.45\textwidth}
    \includegraphics[width=75mm]{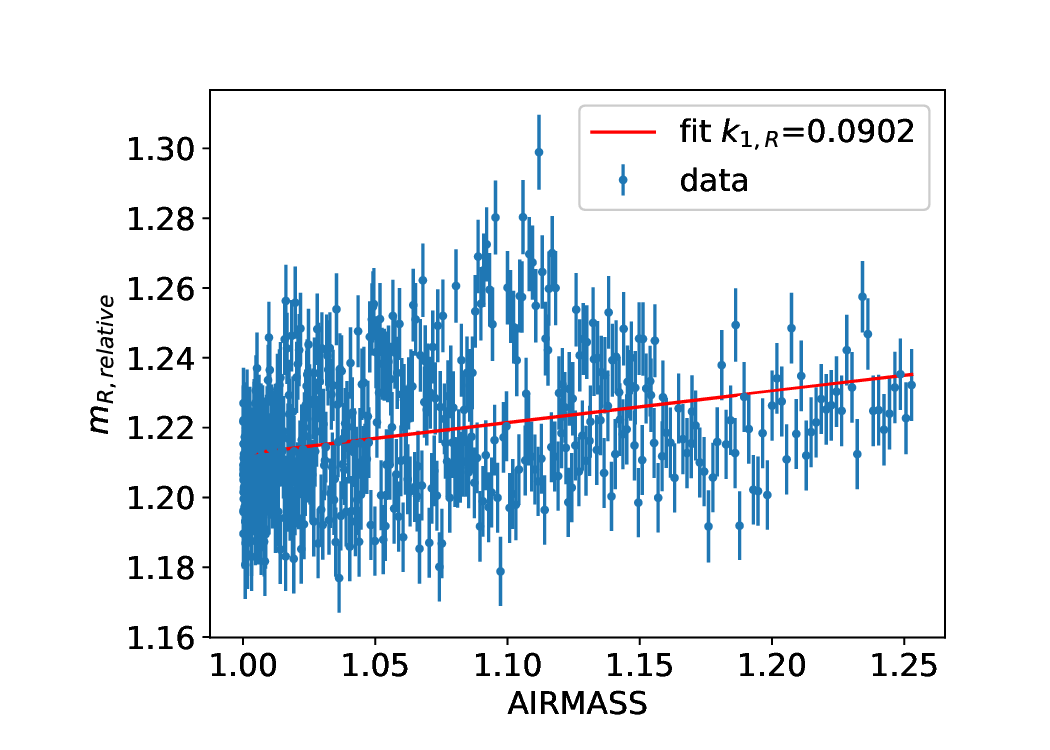}
    \caption{\label{fig:hd165434-R}{\small R band}}
  \end{subfigure}
  \hfill
  \begin{subfigure}[t]{0.45\textwidth}
    \includegraphics[width=75mm]{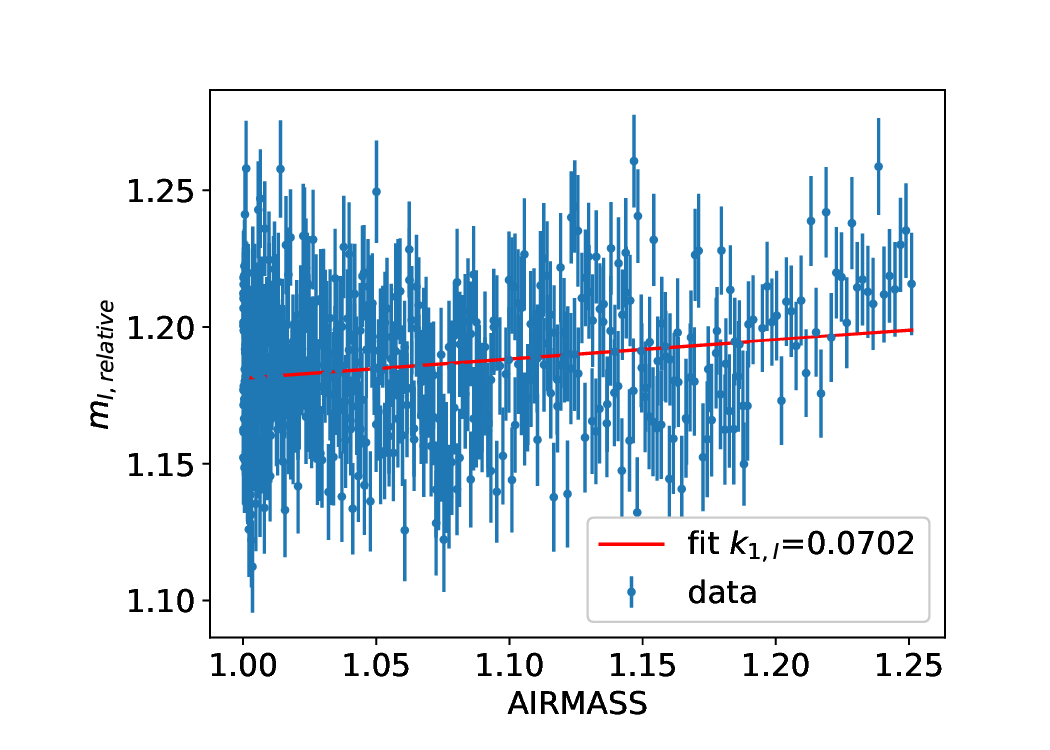}
    \caption{\label{fig:hd165434-I}{\small I band}}
  \end{subfigure}
  \caption{Image of the four bands of HD 165434 taken by our observing system,The blue dots represent the observed data, while the red line represents the fitted line. The x-axis represents airmass, and the y-axis represents the difference between the instrumental magnitude and the GAIA-SP reference magnitude($m_{residual}$).}
  \label{fig:aivsm}
\end{figure}

Figure \ref{fig:aivsm} show the airmass versus magnitude of the standard star HD 165434 in each band. The blue dots represent the observed data, while the red line represents the fitted line. 
The x-axis represents airmass, and the y-axis represents the difference between the instrumental magnitude and the GAIA-SP reference magnitude $m_{residual}$. 
Because the observations were made in summer, the range of airmass values shown here is relatively small,
Table \ref{tab:f_bvri_r} provides a summary of the fitting results in four bands, with $k_1$ and $c_0$ corresponding to the respective coefficients in Equation \ref{eq:st}, $\sigma_{k_1}$ and $\sigma_{c_0}$ represent the simulated errors.The $k_{1,B}$ is $0.335 \pm 0.022$, the $k_{1,V}$ is $0.193 \pm 0.018$, the $k_{1,R}$ is $0.090 \pm 0.013 $, and the $k_{1,I}$ is $0.070 \pm 0.017 $.
\begin{table}[htbp]
	\center
	\caption{The results of the four bands on HD 165434 are obtained by iterative correction} \label{tab:f_bvri_r}
	\begin{tabular}{lcccc}
	\hline\noalign{\smallskip}
	{} & {$k_1$} & {$\sigma_{k_1}$} & {$c_0$} & {$\sigma_{c_0}$} \\
	\hline\noalign{\smallskip}
	B & 0.335 & 0.022 & 0.786 & 0.024 \\
	V & 0.193 & 0.018 & 1.107 & 0.019  \\
	R & 0.090 & 0.013 & 1.122 & 0.014  \\
	I & 0.070 & 0.017 & 1.011 & 0.018 \\
	\noalign{\smallskip}\hline
	\end{tabular}
\end{table}
\\\\
\textit{SA 20-43}

\begin{figure}
  \centering
   \includegraphics[width=100mm]{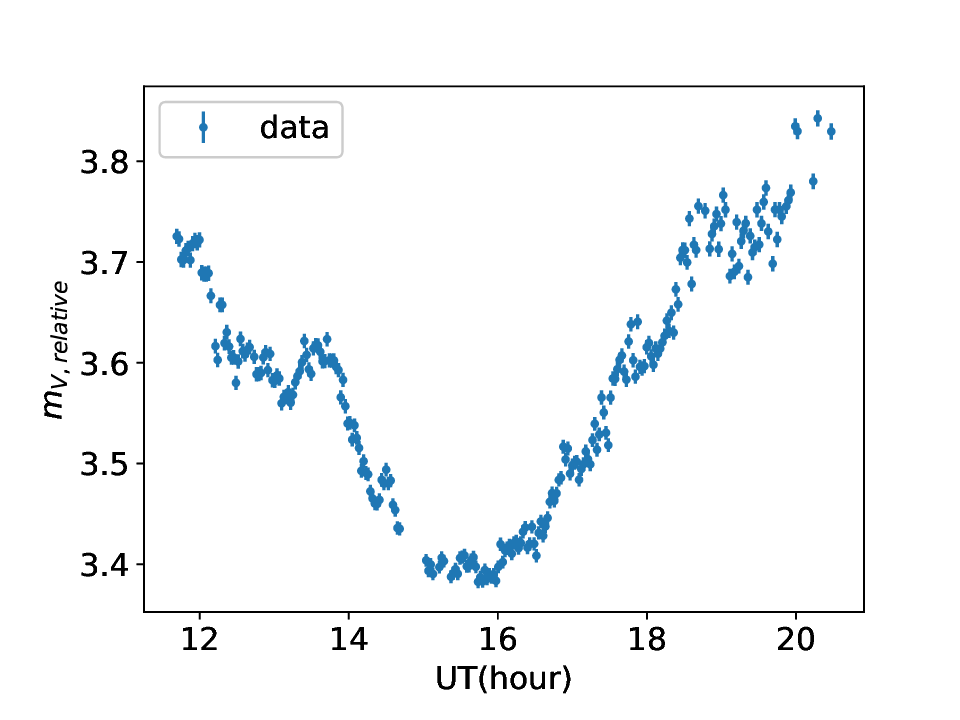}
	 \caption{\label{fig:fol_resi}{\small SA 20-43 V-band $m_{residuals}$ versus UT}}
\end{figure}

\begin{figure}[htbp]
  \centering
   \includegraphics[width=100mm]{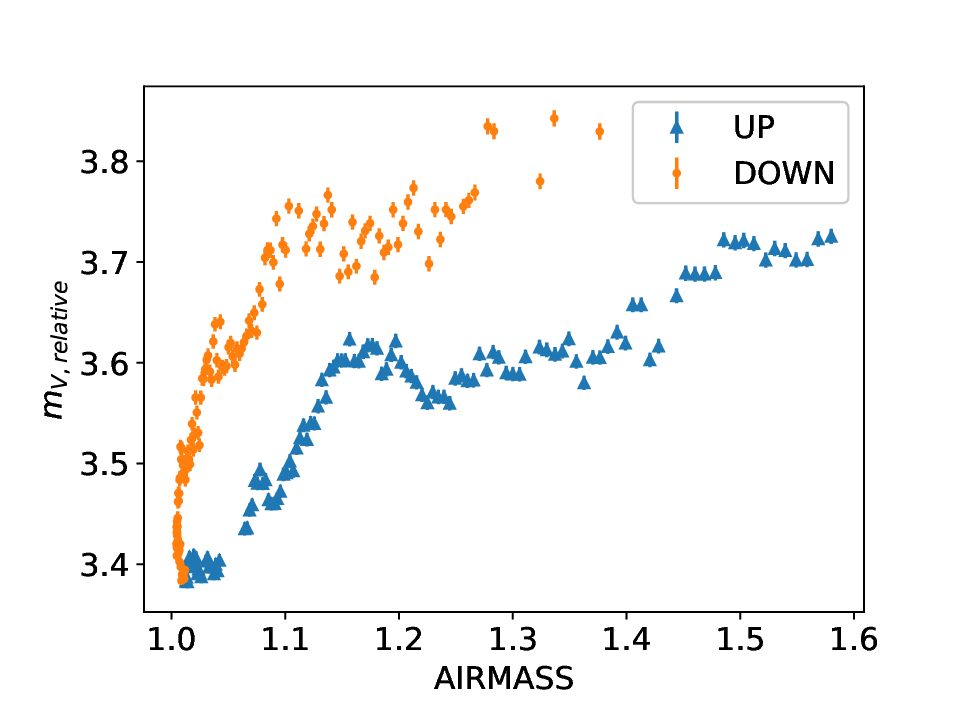}
	  \caption{\label{fig:magair}{\small V-band airmass versus magnitude plot of SA 20-43. Triangles are UP segments, dots are DOWN segments.} }
\end{figure}

 Figure \ref{fig:fol_resi} shows the variation of the standard star's magnitude with respect to the observation time. The x-axis represents the time in UT (hours), while the y-axis represents the difference between the instrumental magnitude and the Gaia-SP magnitude ($m_{residual}$). 
 The variation pattern shows a gradual dimming followed by a sudden brightening, continuing to increase in brightness until reaching the maximum. 
 This point indicates the target reaching the zenith. Subsequently, the target undergoes a descending phase with a continuous dimming, followed by a slight brightening and eventually fading. 
 This pattern is observed for all targets within the field of view, indicating the absence of any variability during the observation time. 
 Figure \ref{fig:magair} displays the airmass vs. magnitude, with the x-axis representing airmass and the y-axis representing the difference between the instrumental magnitude and the Gaia-SP reference magnitude ($m_{residual}$). 
 The blue triangles represent the ascending phase, while the orange dots represent the descending phase.

This problem has also been discussed in the literature(\citealt{Yan+etal+2000},\citealt{Zhou+etal+2001}), and two different methods for its correction: One is to correct for the effect of atmospheric extinction by including a band-independent but time-varying zero correction term in its calibration, which was later replaced by a time-varying atmospheric extinction coefficient term(\citealt{Yan+etal+2000}). 
The other is to use a time-varying atmospheric extinction coefficient term to correct for the effect of atmospheric extinction variations(\citealt{Zhou+etal+2001}). 
In this work, we used the second method to set up a time-dependent term for the atmospheric extinction coefficient. 
Since we're only using one target, there is no the color item. 
To better correct atmospheric extinction, we add a time-varying atmospheric extinction coefficient term. 
For the Equation \ref{eq:st}, based on the corrected relationship of the individual standard stars above, becomes this:
\begin{equation}
    k'_1 = k_1+f(UT) \label{eq:crr_o}
\end{equation}
\begin{equation}
    f(UT)=a\cdot UT+b\label{eq:crr_o1}
\end{equation}
\begin{equation}
    \begin{aligned}
        m_{B,inst}=m_{B,Gaia-sp}+c_{B}+(k_{1,B}+a_{B}\cdot UT+b_{B})\cdot X + constant\\
		m_{V,inst}=m_{V,Gaia-sp}+c_{V}+(k_{1,V}+a_{V}\cdot UT+b_{V})\cdot X + constant\\
		m_{R,inst}=m_{R,Gaia-sp}+c_{R}+(k_{1,R}+a_{R}\cdot UT+b_{R})\cdot X + constant\\
		m_{I,inst}=m_{I,Gaia-sp}+c_{I}+(k_{1,I}+a_{I}\cdot UT+b_{I})\cdot X + constant
    \end{aligned}
    \label{eq:crr2}
\end{equation}

\begin{figure}[htbp]
  \centering
  \begin{subfigure}[t]{0.45\linewidth}
  \centering
   \includegraphics[width=75mm]{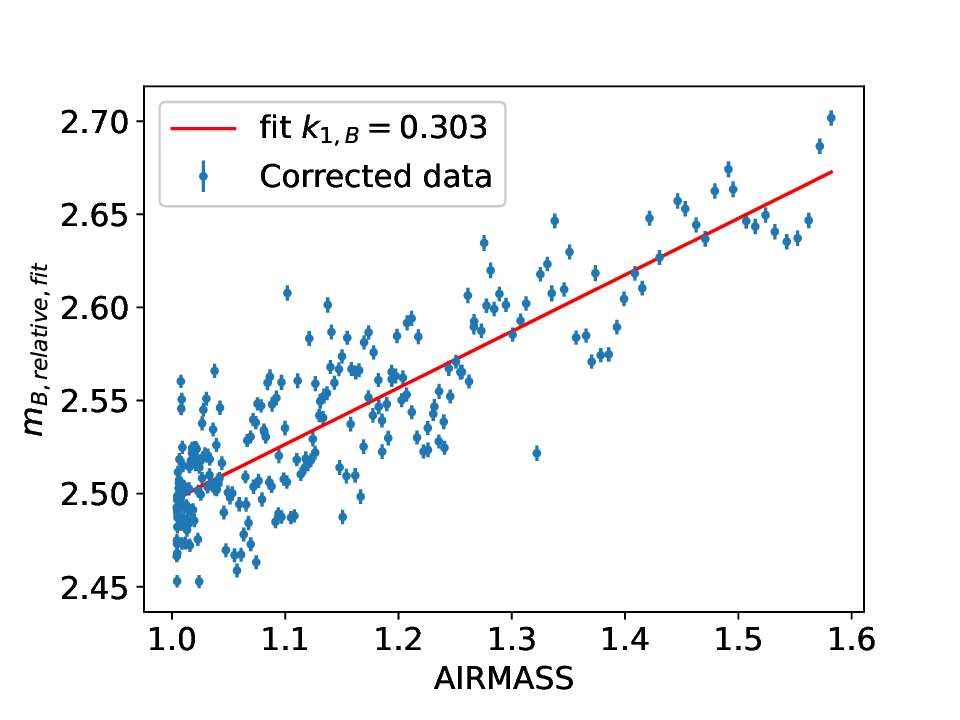}
	  \caption{\label{fig:one-B}{\small B} }
  \end{subfigure}
  \hfill
  \begin{subfigure}[t]{0.45\textwidth}
  \centering
   \includegraphics[width=75mm]{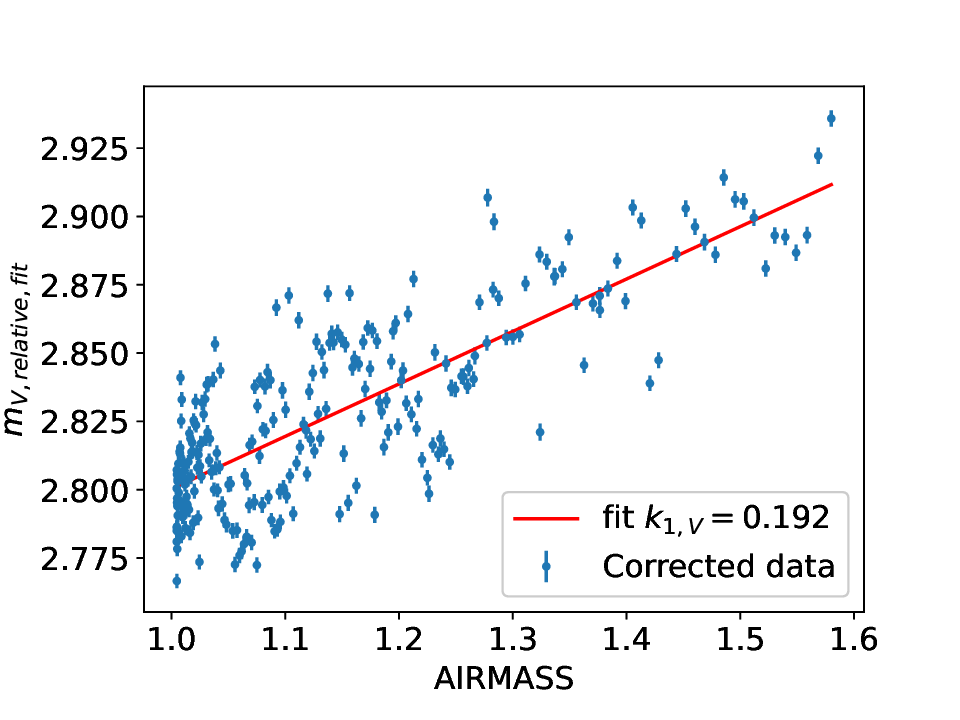}
	  \caption{\label{fig:one-V}{\small V}}
    \end{subfigure}
    \vfill
  \begin{subfigure}[t]{0.45\textwidth}
    \includegraphics[width=75mm]{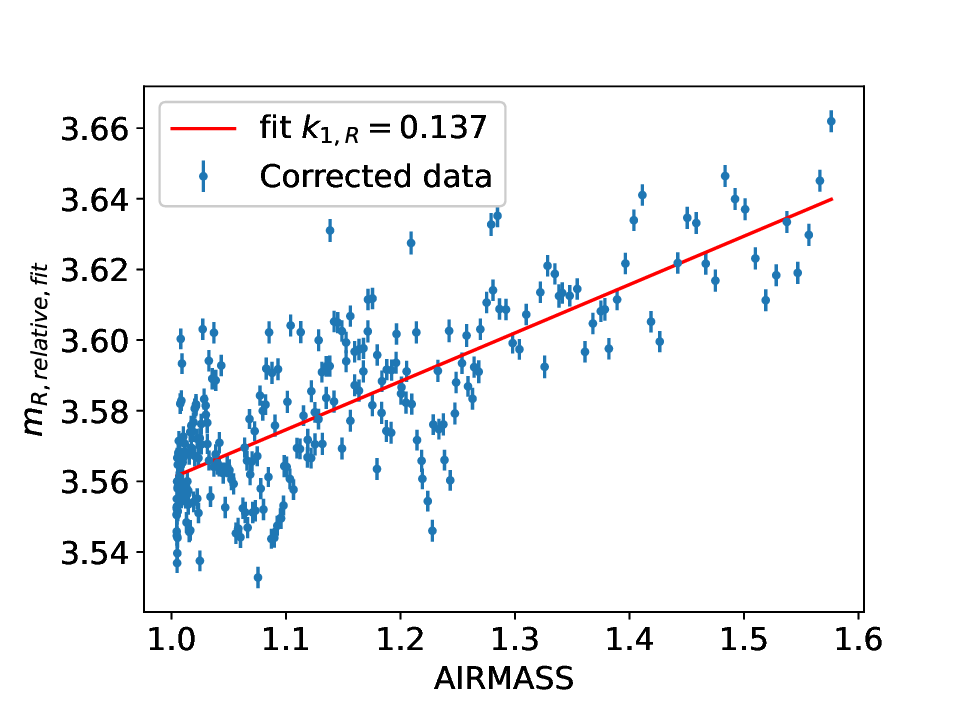}
    \caption{\label{fig:one-R}{\small R}}
  \end{subfigure}
  \hfill
  \begin{subfigure}[t]{0.45\textwidth}
    \includegraphics[width=75mm]{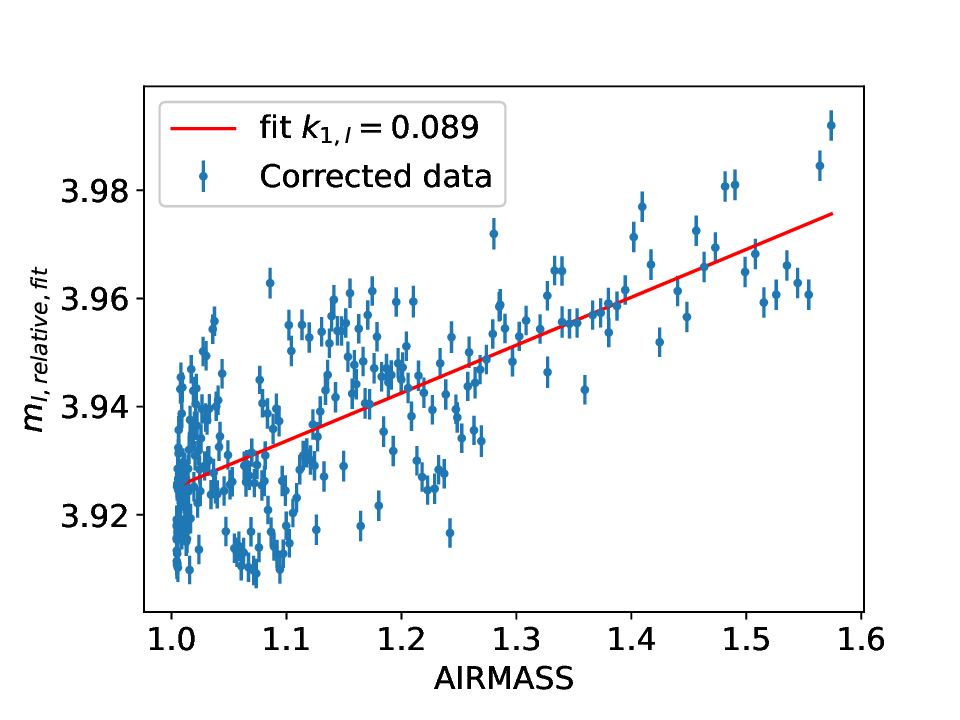}
    \caption{\label{fig:one-I}{\small I}}
  \end{subfigure}
  \caption{The corrected data and fitting results for SA 20-43 are presented. The x-axis represents airmass, while the y-axis represents the difference between the instrumental magnitude and the GAIA-SP reference magnitude ($m_{residual}$). The blue dots represent the corrected data, while the red line represents the linear fit obtained from the corrected data.} \label{fig:iter_air}
\end{figure}

Here, in reality, Equation \ref{eq:crr_o1} is a polynomial. 
We start with the simplest case for our analysis. By fitting the data using this approach, we can obtain the coefficients. 
Subsequently, through a stepwise iteration, we can transform the monomials into a polynomial, thereby fully correcting the observed data. 
This results shown in Figure \ref{fig:iter_air}, where the x-axis represents airmass and the y-axis represents the difference between the instrumental magnitude and the Gaia-sp-SP reference magnitude ($m_{residual}$). 
The blue dots represent the corrected data points, while the red line represents the linear fit of the corrected data. 
The final results are summarized in Table \ref{tab:f_bvri}. The coefficients $k_1$ and $c_0$ in Table \ref{tab:f_bvri} correspond to the respective coefficients in Equation \ref{eq:crr2}. $\sigma_{k_1}$ and $\sigma_{c_0}$ represent the simulated errors. The $k_{1,B}$ is $0.303 \pm 0.013$, the $k_{1,V}$ is $ 0.192 \pm 0.012$, the $k_{1,R}$ is $0.137 \pm 0.009$, and the $k_{1,I}$ is $0.089 \pm 0.006$.
\begin{table}[htbp]
	\center
	\caption{The results of the four bands on SA 20-43 are obtained by iterative correction} \label{tab:f_bvri}
	\begin{tabular}{ccccc}
	\hline\noalign{\smallskip}
	{filter} & {$k_1$} & {$\sigma_{k_1}$} & {$c_0$} & {$\sigma_{c_0}$} \\
	\hline\noalign{\smallskip}
	B & 0.303 & 0.013 & 2.192 & 0.036  \\
	V & 0.192 & 0.012 & 2.622 & 0.032 \\
	R & 0.137 & 0.009 & 3.245 & 0.024 \\
	I & 0.089 & 0.006 & 3.826 & 0.028 \\
	\noalign{\smallskip}\hline
	\end{tabular}
\end{table}
\\\\
\textit{Comparison}

\begin{figure}[htbp]
    \centering
    \includegraphics[width=100mm, angle=0]{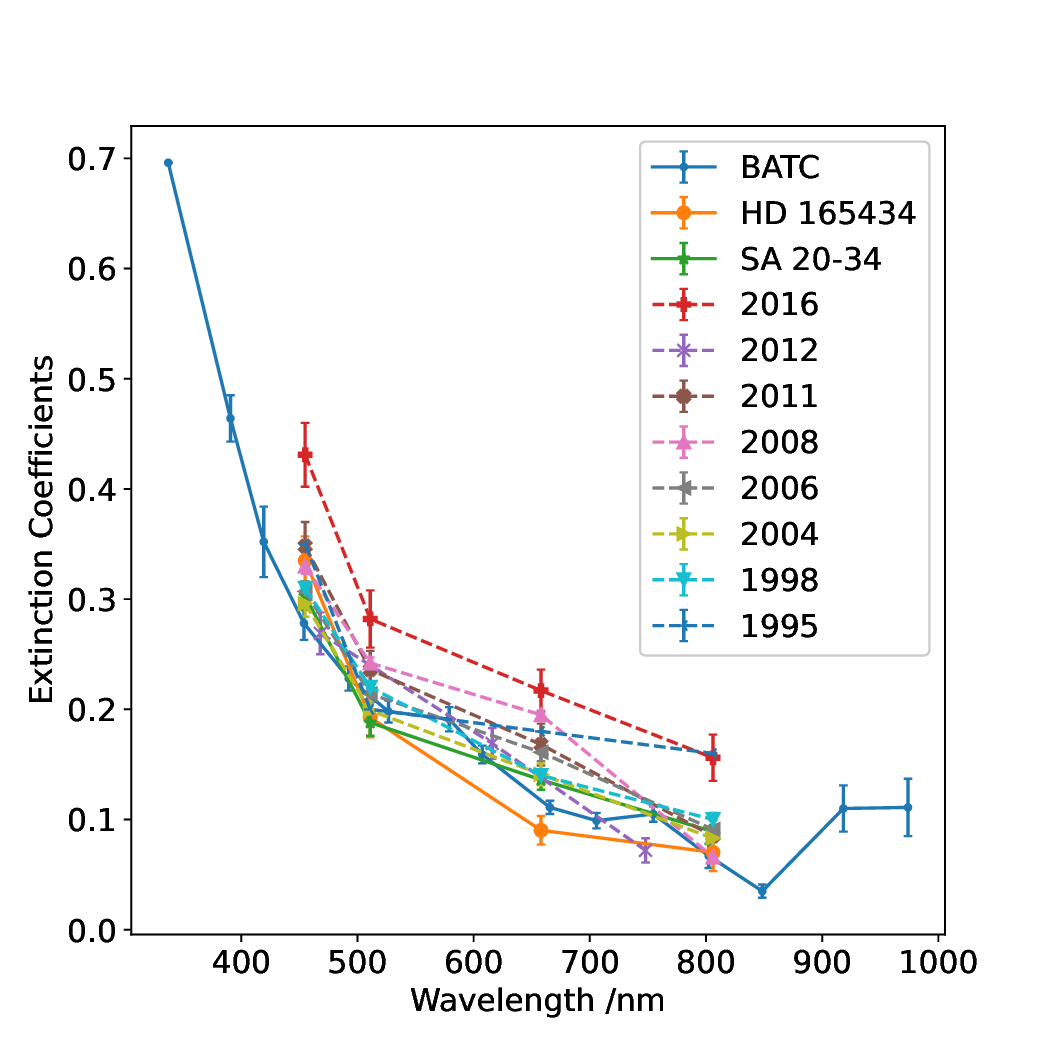}
    \caption{With other working extinction coefficient, The x-axis represents wavelength, while the y-axis represents the extinction coefficient. The data labeled as BATC is sourced from \citealt{Zhou+etal+2001}, the data labeled as 2012 is sourced from \citealt{Zhao+etal+2020}, and the data labeled as 2011 is sourced from reference \citealt{Huang+etal+2012}.}
    \label{fig:filt2}
\end{figure}

The table \ref{tab:ref2} and Figure \ref{fig:filt2} compare all our measured extinction coefficients with those calculated for other telescopes at the Xinglong Observatory(NAOC).\\
\begin{table}[htbp]
	\begin{center}
		\caption{With other working extinction coefficient at the Xinglong Observatory (NAOC)}	\label{tab:ref2}
		\begin{tabular}{ccccccc}
		\hline\noalign{\smallskip}
		{Article} & {Telescope} & {Year} & {$k_{1,B}$} & {$k_{1,V}$} &{$k_{1,R}$} & {$k_{1,I}$} \\
		\hline\noalign{\smallskip}
		    HD 165434 (this article) & 60 cm & 2023 & $0.335 \pm 0.022 $ & $0.193 \pm 0.018 $ & $0.090 \pm 0.013 $ & $0.070 \pm 0.017 $  \\
			SA 20-43 (this article) & 60 cm & 2022 & $0.303 \pm 0.013$ & $ 0.192 \pm 0.012$ & $0.137 \pm 0.009$ & $0.089 \pm 0.006$ \\
			\citealt{Bai+etal+2018} & 85 cm & 2016 & $0.431 \pm 0.029$ & $0.282\pm 0.026$ & $0.217 \pm 0.019$ & $0.156 \pm 0.021$ \\
			\citealt{Huang+etal+2012} & 80 cm & 2011-2012 & $0.348 \pm 0.022$ & $0.236 \pm 0.017 $ & $0.168 \pm 0.019$ & $0.085 \pm 0.021$ \\
			\citealt{Huang+etal+2012} & 80 cm & 2006-2007 & $0.307 \pm 0.009 $ & $0.214\pm 0.008$ & $0.161\pm 0.008$ & $0.091\pm 0.008$ \\
			\citealt{Huang+etal+2012} & 80 cm & 2004-2005 & $0.296 \pm 0.012 $ & $0.199\pm 0.009$ & $0.141\pm 0.010$ & $0.083\pm 0.009$ \\
			\citealt{Zhou+etal+2009} & 85 cm & 2008 & $0.330 \pm 0.007$ & $0.242\pm 0.005 $ & $0.195\pm 0.004$ & $0.066\pm 0.003$  \\
			\citealt{Shi+etal+1998} & 60 cm & 1998 & 0.31 & 0.22 & 0.14 & 0.10 \\
			\citealt{Shi+etal+1998} & 60 cm & 1995 & 0.35 & 0.20 & 0.18 & 0.16 \\
		\noalign{\smallskip}\hline
		\end{tabular}
	\end{center}
\end{table}
The results were nearly identical to others measured at Xinglong Observatory(NAOC).

\subsection{Exoplanet}
\hspace{2em}For exoplanets, we use differential photometry, which compares the light curves of stars within an image's field of view, eliminating the influence of instruments and weather. 
The real used to correct the star, the reference star, the brightness of reference star can not be too different from the target star, the distance between the figure and the target can not be too far. Either the sigma of the light curve can not be too large, i.e. 0.01, to ensure that it is not a variable star.
The number of reference star, so that in the image is the most appropriate. 
A comparable star, with the same requirements as the reference star, is also required at the end for final comparison.
\\\\
\textit{HAT-P-32 b}

\begin{figure}[htbp]
    \centering
    \includegraphics[width=100mm, angle=0]{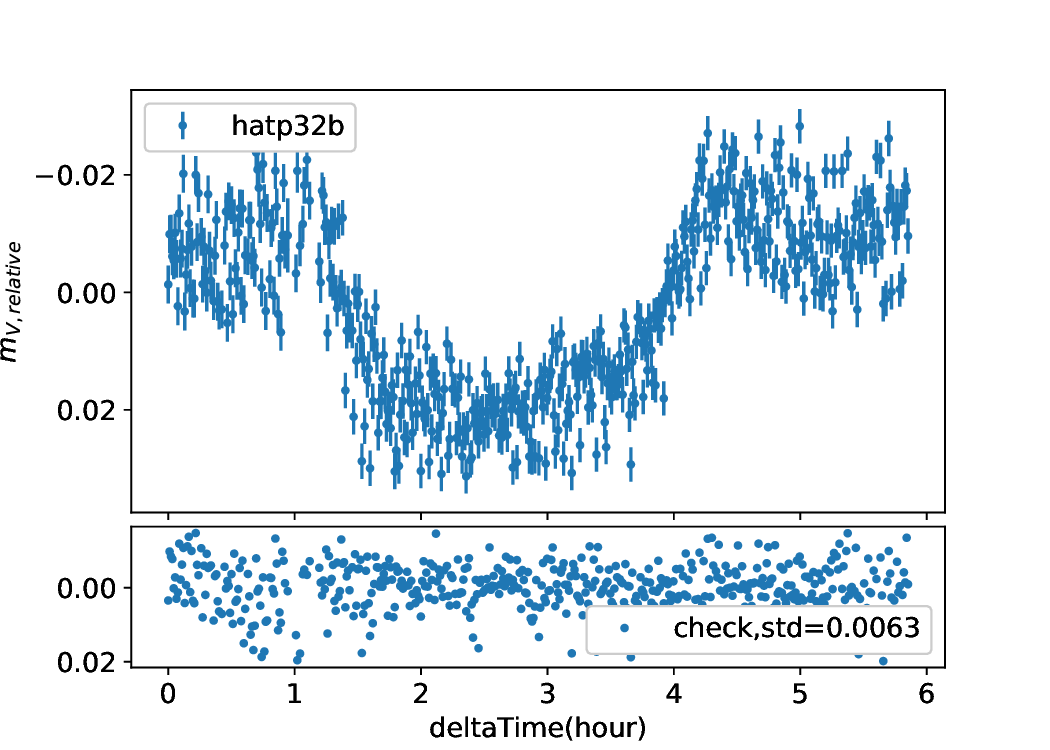}
    \caption{ HAT-P-32 b original data. The upper panel shows the observed eclipse light curve of HAT-P-32 b, while the lower panel displays the photometric variation of the comparison stars. The same x-axis is deltaTime (hour), representing the time span from the beginning to the end of the observation. The y-axis of both panels represents the relative magnitude, $m_{V,relative}$.}
    \label{fig:hatp-o}
\end{figure}

\begin{figure}[htbp]
  \begin{subfigure}[t]{0.45\linewidth}
  \centering
   \includegraphics[width=75mm]{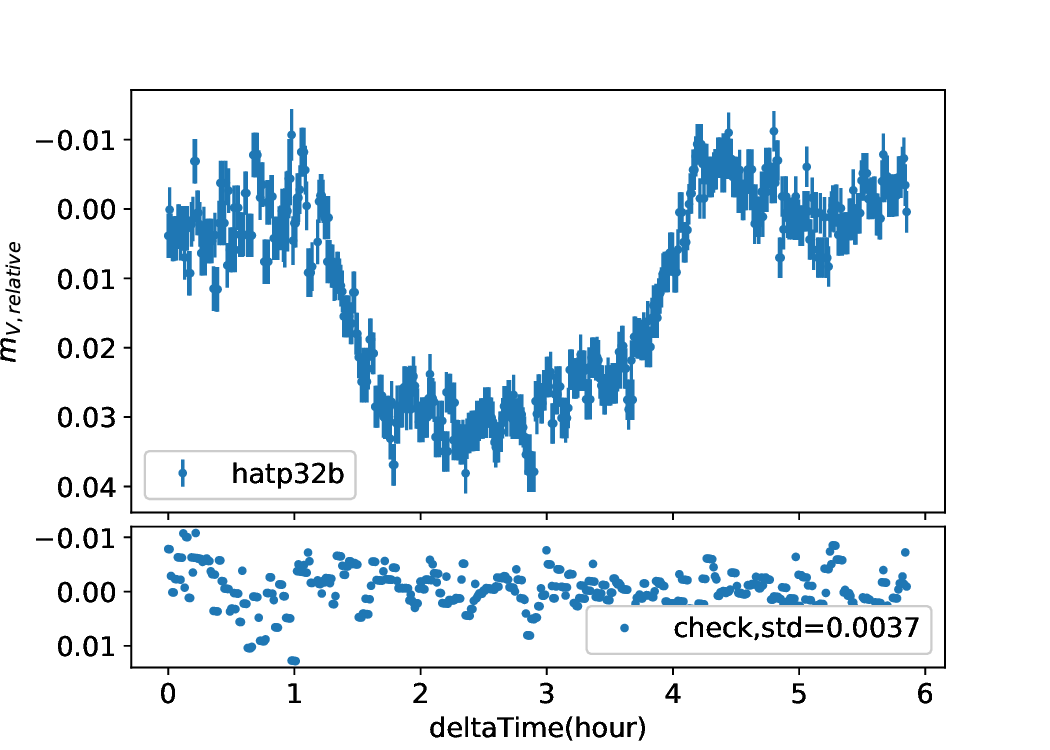}
	  \caption{\label{fig:hatp}{\small The eclipse light curve of HAT-P-32 b is shown in Figure \ref{fig:hatp-o}, with data binned at a size of 5.} }
  \end{subfigure}%
  \hfill
  \begin{subfigure}[t]{0.45\textwidth}
  \centering
   \includegraphics[width=75mm]{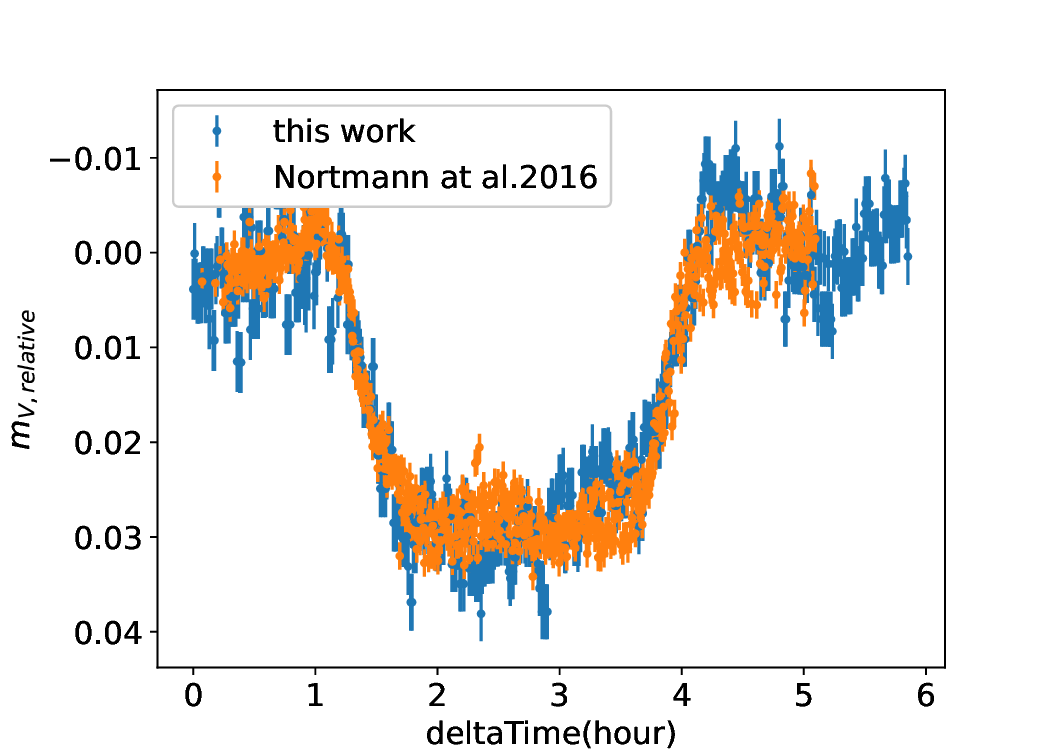}
	  \caption{\label{fig:hatp_c}{\small Comparison of eclipse light curves of HAT-P-32 b. Blue data points represent observations from this observation, i.e. the data in Fig. 1, and orange data points represent eclipse light curves from \citealt{Nortmann+etal+2016}.}}
  \end{subfigure}%
  \caption{HATP-32b light curve and HATP-32b light curve compare} \label{fig:explant1}
\end{figure}

For differential photometry, we can obtain its light curve. 
Figure \ref{fig:hatp-o} is the light curve extracted by us after differential correction. 
The top panel is the light curve of the host star of this exoplanet, and the bottom panel is the light curve of the reference star. The std in label in the bottom panel stands for differential photometric accuracy, which the scatter is 0.0063mag. Then, we combine the five frames into one frame (binned 5) to increase the SNR,in the Figure  \ref{fig:hatp}. 
In the Figure \ref{fig:hatp} represents our photographic acquisition of HAT-P-32 b and its control star, and we can see the obvious eclipse light variation in the target, which the scatter is 0.0037mag; 
Figure \ref{fig:hatp_c} is the comparison of our light curve with those by other work from literature, and they completely
agree each other, but the basic trend is clear. 
 The data used in Figure \ref{fig:hatp_c}.
 The observing equipment they used in the article (\citealt{Nortmann+etal+2016,Hartman+etal+2011}) was the OSIRIS instrument of the Gran Telescopio CANARIAS (GTC) in long-slit spectral mode.
\\\\
\textit{WASP-33 b}

\begin{figure}[htbp]
    \centering
    \includegraphics[width=100mm, angle=0]{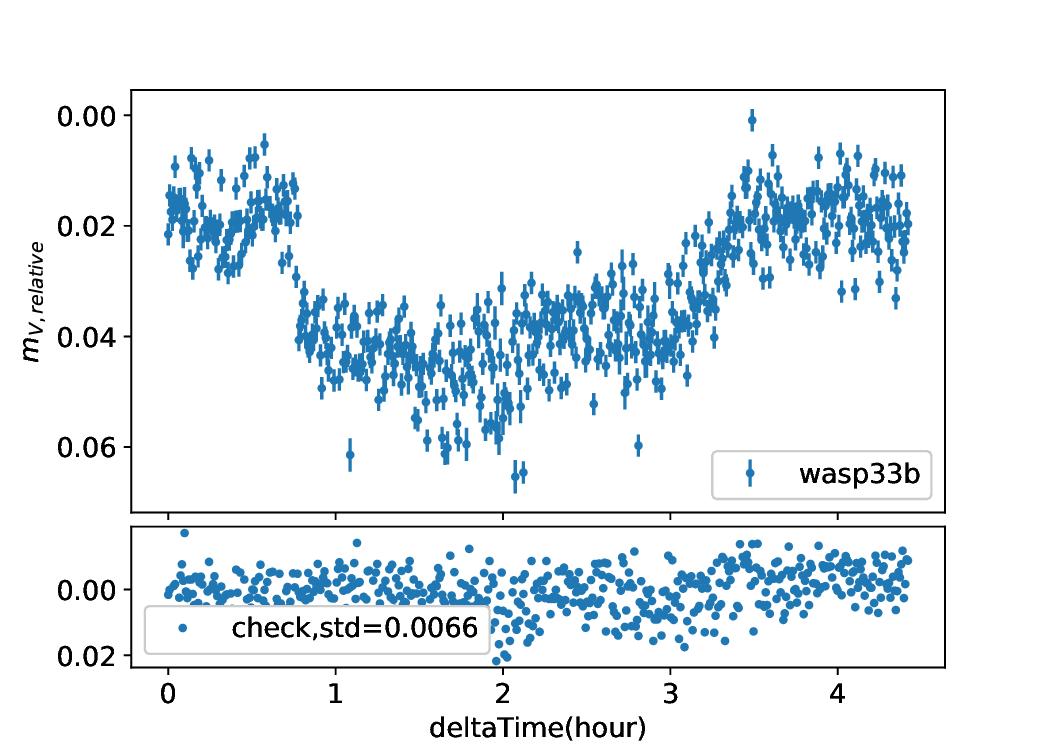}
    \caption{The upper panel shows the observed eclipse light curve of WASP-33 b, while the lower panel displays the variation of the comparison stars. The same x-axis is deltaTime (hour), representing the time span from the beginning to the end of the observation. The y-axis of both panels represents the relative magnitude, $m_{V,relative}$.}
    \label{fig:wasp-o}
\end{figure}

\begin{figure}[htbp]
  \begin{subfigure}[t]{0.45\linewidth}
  \centering
   \includegraphics[width=75mm]{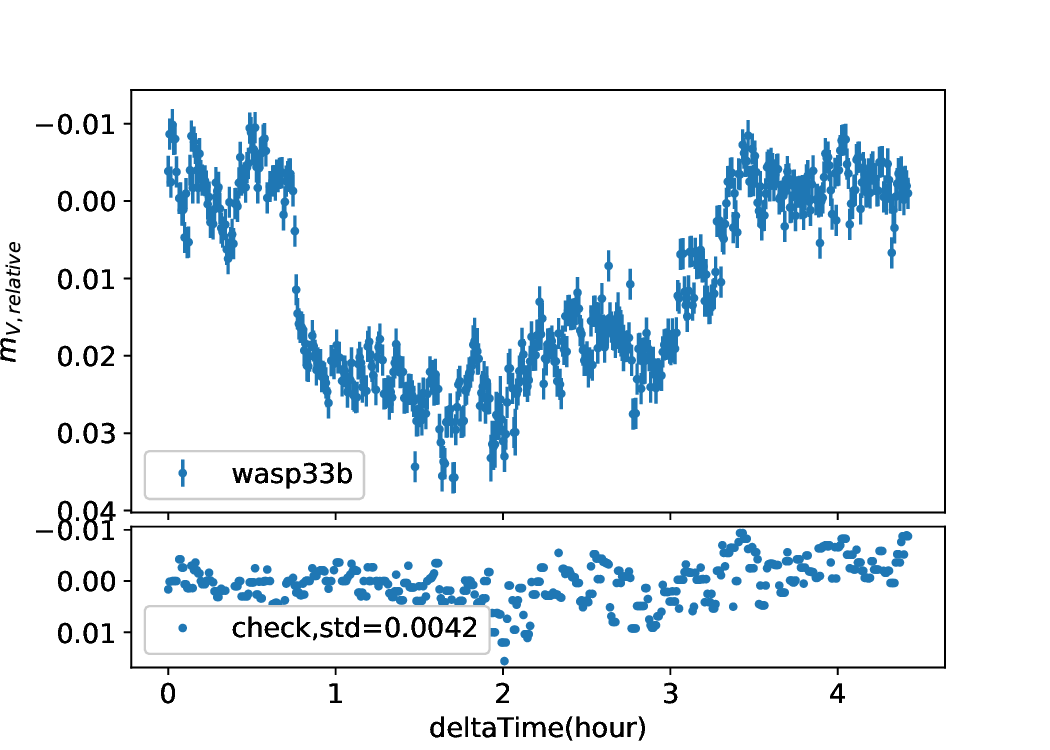}
	  \caption{\label{fig:wasp}{\small The eclipse light curve of WASP-33 b is shown in Figure \ref{fig:wasp-o}, with data binned at a size of 5.} }
  \end{subfigure}%
  \hfill
  \begin{subfigure}[t]{0.45\textwidth}
  \centering
   \includegraphics[width=75mm]{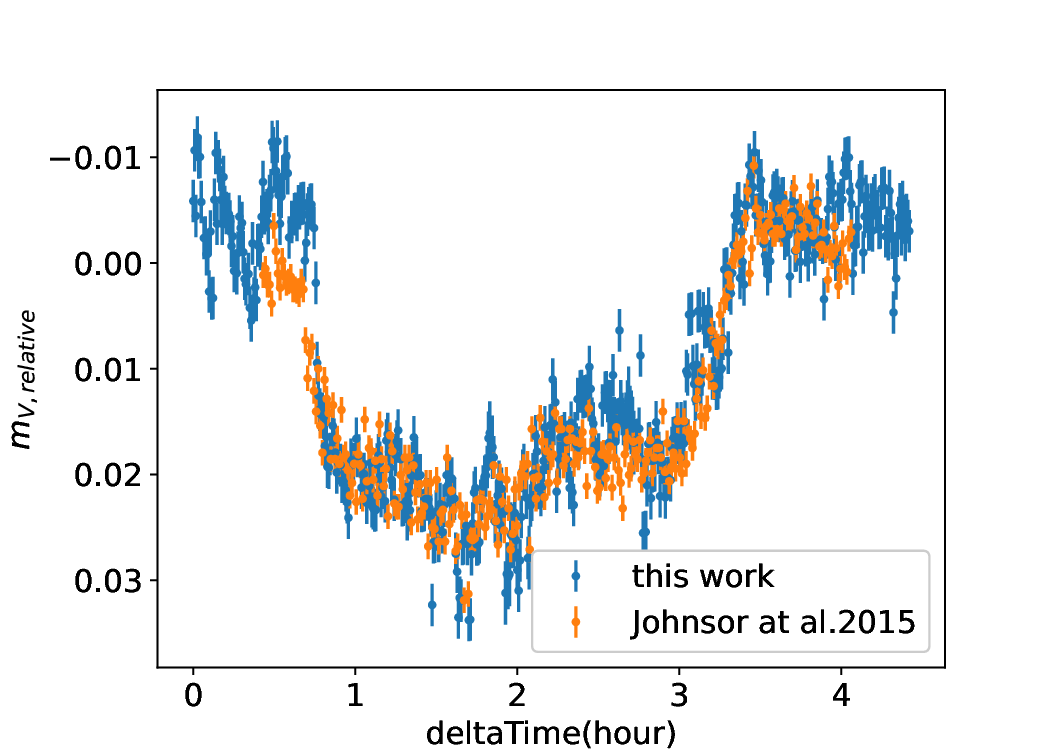}
	  \caption{\label{fig:wasp_c}{\small Comparison of eclipse light curves of WASP-33 b. Blue data points represent observations from this observation, i.e. the data in Fig. 1, and orange data points represent eclipse light curves from \citealt{Johnson+etal+2015}.}}
  \end{subfigure}%
  \caption{WASP-33 b light curve and WASP-33 b light curve compare} \label{fig:explant2}
\end{figure}

Since WASP-33 is a $\delta$ Scuti variable star (\citealt{Herrero+etal+2011}), it means that the light curve of the exoplanet contain the variation of the host star. 
The relationships in Figures \ref{fig:wasp-o} and \ref{fig:wasp} for WASP-33 b are the same as those in Figures \ref{fig:hatp-o} , which the scatter is 0.0066mag, and \ref{fig:hatp} for HAT-P-32 b, which are taken before and after binned 5. Figure \ref{fig:wasp} shows our light curve of WASP-33 b and its host star, even its light variation of 0.02mag we can see clearly, while the precision of the former reference star is about 0.004mag in the best situation. 
This depth of the evolutionary eclipse may include the light variations of the host star.
Compare WASP-33 b with data obtained by other observers using CCD observations, our result also agree well with \citealt{Johnson+etal+2015} obtained from the 0.3-meter telescope at Monte Carbre Observatory and the 0.8-meter telescope at Montserrat Observatory in the Figure \ref{fig:wasp_c}.

Then we compare the observing performance of the 60-cm telescope under CCD. From the article (\citealt{Wang+etal+2017} $\&$ \citealt{Wang+etal+2019} ) we see that they have 60-cm frames set up for exoplanets with differential photometric accuracy around $0.0016-0.0063mag$. And the differential photometric accuracy obtained from our measurements is between $0.003-0.004mag$, which is comparable or even smaller than the references. 
Therefore, we conclude CMOS is suitable for observing exoplanets and can replace the CCD for this program.

\section{Summary and conclusions}
\label{sect:conclusion}
\hspace{2em}In this work, the tests mainly includes these items: Supersky flat field, the accuracy of photometry, the accuracy of differential photometry, and the extinction coefficients with CMOS observations mounted in 60-cm telescope at
the Xinglong Observatory(NAOC) . 
 Based on the tests, flat field correction for CMOS + 60cm telescope
could be better than 1\% at BVRI bands, and thus is the change of flat field every day during the observation period.
Flat field can be replaced by that of another day during our observations.
Therefore, we conclude that the camera's performance is stable.
The following conclusions: The open clusters results in a photometric accuracy of 0.02mag. The calculated extinction coefficient of our work is compared with other coefficients observed at the Xinglong Observatory(NAOC) using the standard star, revealing a significant difference. The differential photometric accuracy of 0.004mag was for exoplanets.

Therefore, we can conclude that the observation results of the observing system in this paper are in accordance with our expectations and can be used in scientific observations. This SONY IMX455 CMOS  sensor is satisfy with our requirements and can replace CCD.\\

In the future, we will test the pixel stability of this camera. We will also test it on other devices with exposure times of 60 seconds, 300 seconds, 600 seconds, and 1200 seconds. Additionally, we will study the dark current at each corresponding exposure time.

\begin{acknowledgements}
Supported by the Strategic Priority Research Program of the Chinese Academy of Sciences (Grant No. XDB0550100, XDB0550000), and supported by National Key R\&D Program of China (Grant No.2023YFA1609700). We also acknowledge the National Natural Science Foundation of China (NSFC No.12090041; 12090040). The science research grants from the China Manned Space Project with NO. CMS-CSST-2021-B03. We also acknowledge the support of the staff at the Xinglong Observatory (NAOC).

\end{acknowledgements}

\label{lastpage}

\end{document}